\documentclass[aps,reprint,superscriptaddress,nofootinbib,amsmath,amssymb,prd]{revtex4-2}

\usepackage{graphicx} 
\usepackage[caption=false]{subfig}
\usepackage{afterpage}
\usepackage{booktabs}
\usepackage[hidelinks,colorlinks=true,linktoc=page,citecolor=blue,filecolor=blue,urlcolor=blue,linkcolor=blue]{hyperref} 
\usepackage{multirow}
\usepackage[normalem]{ulem}
\usepackage{xcolor}
\usepackage{adjustbox}
\usepackage{physics}
\usepackage{url}
\usepackage{orcidlink}
\usepackage{bm}
\usepackage{float}
\usepackage{makecell}

\usepackage{array} 
\usepackage[nameinlink,noabbrev]{cleveref}

\crefname{equation}{Eq.}{Eqs.}
\crefname{section}{\S}{\S}
\crefname{figure}{Figure}{Figures}
\crefname{table}{Table}{Tables}
\crefname{appendix}{Appendix}{Appendices}
\Crefname{figure}{Figure}{Figures}
\Crefname{section}{Section}{Sections}
\Crefname{table}{Table}{Tables}

\newcommand{\fnl}[0]{f_{\rm NL}^{\rm loc}}

\newcommand{\ie}[0]{\textit{i.e.} }
\newcommand{\eg}[0]{\textit{e.g.} }

\newcommand{\seliminline}[1]{{\color{blue}{#1}}}

\begin{document} 
\title{Measurement of the galaxy-velocity power spectrum of DESI tracers with the kinematic Sunyaev–Zeldovich effect using DESI DR2 and ACT DR6.}

\author{Edmond~Chaussidon\orcidlink{0000-0001-8996-4874}}
\email{echaussidon@lbl.gov}
\affiliation{Lawrence Berkeley National Laboratory, 1 Cyclotron Road, Berkeley, CA 94720, USA}

\author{Selim~C.~Hotinli\orcidlink{0000-0003-0061-8188}}
\affiliation{Perimeter Institute for Theoretical Physics, 31 Caroline St. North, Waterloo, ON N2L 2Y5, Canada}

\author{Simone~Ferraro\orcidlink{0000-0003-4992-7854}}
\affiliation{Lawrence Berkeley National Laboratory, 1 Cyclotron Road, Berkeley, CA 94720, USA}
\affiliation{Berkeley Center for Cosmological Physics, Department of Physics,
University of California, Berkeley, CA 94720, USA}

\author{Kendrick~Smith}
\affiliation{Perimeter Institute for Theoretical Physics, 31 Caroline St. North, Waterloo, ON N2L 2Y5, Canada}

\author{Xinyi~Chen\orcidlink{0000-0003-3456-0957}}
\affiliation{Center for Cosmology and AstroParticle Physics, The Ohio State University, 191 West Woodruff Avenue, Columbus, OH 43210, USA}
\affiliation{Department of Physics, The Ohio State University, 191 West Woodruff Avenue, Columbus, OH 43210, USA}

\author{J.~Aguilar}
\affiliation{Lawrence Berkeley National Laboratory, 1 Cyclotron Road, Berkeley, CA 94720, USA}

\author{S.~Ahlen\orcidlink{0000-0001-6098-7247}}
\affiliation{Department of Physics, Boston University, 590 Commonwealth Avenue, Boston, MA 02215 USA}

\author{D.~Bianchi\orcidlink{0000-0001-9712-0006}}
\affiliation{Dipartimento di Fisica ``Aldo Pontremoli'', Universit\`a degli Studi di Milano, Via Celoria 16, I-20133 Milano, Italy}
\affiliation{INAF-Osservatorio Astronomico di Brera, Via Brera 28, 20122 Milano, Italy}

\author{D.~Brooks}
\affiliation{Department of Physics \& Astronomy, University College London, Gower Street, London, WC1E 6BT, UK}

\author{T.~Claybaugh}
\affiliation{Lawrence Berkeley National Laboratory, 1 Cyclotron Road, Berkeley, CA 94720, USA}

\author{A.~Cuceu\orcidlink{0000-0002-2169-0595}}
\affiliation{Lawrence Berkeley National Laboratory, 1 Cyclotron Road, Berkeley, CA 94720, USA}

\author{A.~de la Macorra\orcidlink{0000-0002-1769-1640}}
\affiliation{Instituto de F\'{\i}sica, Universidad Nacional Aut\'{o}noma de M\'{e}xico,  Circuito de la Investigaci\'{o}n Cient\'{\i}fica, Ciudad Universitaria, Cd. de M\'{e}xico  C.~P.~04510,  M\'{e}xico}

\author{B.~Dey\orcidlink{0000-0002-5665-7912}}
\affiliation{Department of Astronomy \& Astrophysics, University of Toronto, Toronto, ON M5S 3H4, Canada}
\affiliation{Department of Physics \& Astronomy and Pittsburgh Particle Physics, Astrophysics, and Cosmology Center (PITT PACC), University of Pittsburgh, 3941 O'Hara Street, Pittsburgh, PA 15260, USA}

\author{P.~Doel}
\affiliation{Department of Physics \& Astronomy, University College London, Gower Street, London, WC1E 6BT, UK}

\author{A.~Font-Ribera\orcidlink{0000-0002-3033-7312}}
\affiliation{Instituci\'{o} Catalana de Recerca i Estudis Avan\c{c}ats, Passeig de Llu\'{\i}s Companys, 23, 08010 Barcelona, Spain}
\affiliation{Institut de F\'{i}sica d’Altes Energies (IFAE), The Barcelona Institute of Science and Technology, Edifici Cn, Campus UAB, 08193, Bellaterra (Barcelona), Spain}

\author{J.~E.~Forero-Romero\orcidlink{0000-0002-2890-3725}}
\affiliation{Departamento de F\'isica, Universidad de los Andes, Cra. 1 No. 18A-10, Edificio Ip, CP 111711, Bogot\'a, Colombia}
\affiliation{Observatorio Astron\'omico, Universidad de los Andes, Cra. 1 No. 18A-10, Edificio H, CP 111711 Bogot\'a, Colombia}

\author{E.~Gaztañaga\orcidlink{0000-0001-9632-0815}}
\affiliation{Institut d'Estudis Espacials de Catalunya (IEEC), c/ Esteve Terradas 1, Edifici RDIT, Campus PMT-UPC, 08860 Castelldefels, Spain}
\affiliation{Institute of Cosmology and Gravitation, University of Portsmouth, Dennis Sciama Building, Portsmouth, PO1 3FX, UK}
\affiliation{Institute of Space Sciences, ICE-CSIC, Campus UAB, Carrer de Can Magrans s/n, 08913 Bellaterra, Barcelona, Spain}

\author{S.~Gontcho A Gontcho\orcidlink{0000-0003-3142-233X}}
\affiliation{University of Virginia, Department of Astronomy, Charlottesville, VA 22904, USA}

\author{G.~Gutierrez}
\affiliation{Fermi National Accelerator Laboratory, PO Box 500, Batavia, IL 60510, USA}

\author{J.~Guy\orcidlink{0000-0001-9822-6793}}
\affiliation{Lawrence Berkeley National Laboratory, 1 Cyclotron Road, Berkeley, CA 94720, USA}

\author{H.~K.~Herrera-Alcantar\orcidlink{0000-0002-9136-9609}}
\affiliation{Institut d'Astrophysique de Paris. 98 bis boulevard Arago. 75014 Paris, France}
\affiliation{IRFU, CEA, Universit\'{e} Paris-Saclay, F-91191 Gif-sur-Yvette, France}

\author{K.~Honscheid\orcidlink{0000-0002-6550-2023}}
\affiliation{Center for Cosmology and AstroParticle Physics, The Ohio State University, 191 West Woodruff Avenue, Columbus, OH 43210, USA}
\affiliation{Department of Physics, The Ohio State University, 191 West Woodruff Avenue, Columbus, OH 43210, USA}

\author{C.~Howlett\orcidlink{0000-0002-1081-9410}}
\affiliation{School of Mathematics and Physics, University of Queensland, Brisbane, QLD 4072, Australia}

\author{D.~Huterer\orcidlink{0000-0001-6558-0112}}
\affiliation{Department of Physics, University of Michigan, 450 Church Street, Ann Arbor, MI 48109, USA}
\affiliation{University of Michigan, 500 S. State Street, Ann Arbor, MI 48109, USA}

\author{M.~Ishak\orcidlink{0000-0002-6024-466X}}
\affiliation{Department of Physics, The University of Texas at Dallas, 800 W. Campbell Rd., Richardson, TX 75080, USA}

\author{R.~Joyce\orcidlink{0000-0003-0201-5241}}
\affiliation{NSF NOIRLab, 950 N. Cherry Ave., Tucson, AZ 85719, USA}

\author{D.~Kirkby\orcidlink{0000-0002-8828-5463}}
\affiliation{Department of Physics and Astronomy, University of California, Irvine, 92697, USA}

\author{A.~Kremin\orcidlink{0000-0001-6356-7424}}
\affiliation{Lawrence Berkeley National Laboratory, 1 Cyclotron Road, Berkeley, CA 94720, USA}

\author{O.~Lahav\orcidlink{0000-0002-1134-9035}}
\affiliation{Department of Physics \& Astronomy, University College London, Gower Street, London, WC1E 6BT, UK}

\author{M.~Landriau\orcidlink{0000-0003-1838-8528}}
\affiliation{Lawrence Berkeley National Laboratory, 1 Cyclotron Road, Berkeley, CA 94720, USA}

\author{L.~Le~Guillou\orcidlink{0000-0001-7178-8868}}
\affiliation{Sorbonne Universit\'{e}, CNRS/IN2P3, Laboratoire de Physique Nucl\'{e}aire et de Hautes Energies (LPNHE), FR-75005 Paris, France}

\author{M.~Manera\orcidlink{0000-0003-4962-8934}}
\affiliation{Departament de F\'{i}sica, Serra H\'{u}nter, Universitat Aut\`{o}noma de Barcelona, 08193 Bellaterra (Barcelona), Spain}
\affiliation{Institut de F\'{i}sica d’Altes Energies (IFAE), The Barcelona Institute of Science and Technology, Edifici Cn, Campus UAB, 08193, Bellaterra (Barcelona), Spain}

\author{A.~Meisner\orcidlink{0000-0002-1125-7384}}
\affiliation{NSF NOIRLab, 950 N. Cherry Ave., Tucson, AZ 85719, USA}

\author{R.~Miquel}
\affiliation{Instituci\'{o} Catalana de Recerca i Estudis Avan\c{c}ats, Passeig de Llu\'{\i}s Companys, 23, 08010 Barcelona, Spain}
\affiliation{Institut de F\'{i}sica d’Altes Energies (IFAE), The Barcelona Institute of Science and Technology, Edifici Cn, Campus UAB, 08193, Bellaterra (Barcelona), Spain}

\author{S.~Nadathur\orcidlink{0000-0001-9070-3102}}
\affiliation{Institute of Cosmology and Gravitation, University of Portsmouth, Dennis Sciama Building, Portsmouth, PO1 3FX, UK}

\author{J.~A.~Newman\orcidlink{0000-0001-8684-2222}}
\affiliation{Department of Physics \& Astronomy and Pittsburgh Particle Physics, Astrophysics, and Cosmology Center (PITT PACC), University of Pittsburgh, 3941 O'Hara Street, Pittsburgh, PA 15260, USA}

\author{N.~Palanque-Delabrouille\orcidlink{0000-0003-3188-784X}}
\affiliation{IRFU, CEA, Universit\'{e} Paris-Saclay, F-91191 Gif-sur-Yvette, France}
\affiliation{Lawrence Berkeley National Laboratory, 1 Cyclotron Road, Berkeley, CA 94720, USA}

\author{W.~J.~Percival\orcidlink{0000-0002-0644-5727}}
\affiliation{Department of Physics and Astronomy, University of Waterloo, 200 University Ave W, Waterloo, ON N2L 3G1, Canada}
\affiliation{Perimeter Institute for Theoretical Physics, 31 Caroline St. North, Waterloo, ON N2L 2Y5, Canada}
\affiliation{Waterloo Centre for Astrophysics, University of Waterloo, 200 University Ave W, Waterloo, ON N2L 3G1, Canada}

\author{F.~Prada\orcidlink{0000-0001-7145-8674}}
\affiliation{Instituto de Astrof\'{i}sica de Andaluc\'{i}a (CSIC), Glorieta de la Astronom\'{i}a, s/n, E-18008 Granada, Spain}

\author{I.~P\'erez-R\`afols\orcidlink{0000-0001-6979-0125}}
\affiliation{Departament de F\'isica, EEBE, Universitat Polit\`ecnica de Catalunya, c/Eduard Maristany 10, 08930 Barcelona, Spain}

\author{G.~Rossi}
\affiliation{Department of Physics and Astronomy, Sejong University, 209 Neungdong-ro, Gwangjin-gu, Seoul 05006, Republic of Korea}

\author{L.~Samushia\orcidlink{0000-0002-1609-5687}}
\affiliation{Abastumani Astrophysical Observatory, Tbilisi, GE-0179, Georgia}
\affiliation{Department of Physics, Kansas State University, 116 Cardwell Hall, Manhattan, KS 66506, USA}
\affiliation{Faculty of Natural Sciences and Medicine, Ilia State University, 0194 Tbilisi, Georgia}

\author{E.~Sanchez\orcidlink{0000-0002-9646-8198}}
\affiliation{CIEMAT, Avenida Complutense 40, E-28040 Madrid, Spain}

\author{D.~Schlegel}
\affiliation{Lawrence Berkeley National Laboratory, 1 Cyclotron Road, Berkeley, CA 94720, USA}

\author{M.~Schubnell}
\affiliation{Department of Physics, University of Michigan, 450 Church Street, Ann Arbor, MI 48109, USA}
\affiliation{University of Michigan, 500 S. State Street, Ann Arbor, MI 48109, USA}

\author{H.~Seo\orcidlink{0000-0002-6588-3508}}
\affiliation{Department of Physics \& Astronomy, Ohio University, 139 University Terrace, Athens, OH 45701, USA}

\author{J.~Silber\orcidlink{0000-0002-3461-0320}}
\affiliation{Lawrence Berkeley National Laboratory, 1 Cyclotron Road, Berkeley, CA 94720, USA}

\author{D.~Sprayberry}
\affiliation{NSF NOIRLab, 950 N. Cherry Ave., Tucson, AZ 85719, USA}

\author{G.~Tarl\'{e}\orcidlink{0000-0003-1704-0781}}
\affiliation{University of Michigan, 500 S. State Street, Ann Arbor, MI 48109, USA}

\author{B.~A.~Weaver}
\affiliation{NSF NOIRLab, 950 N. Cherry Ave., Tucson, AZ 85719, USA}

\author{C.~Y\`eche\orcidlink{0000-0001-5146-8533}}
\affiliation{IRFU, CEA, Universit\'{e} Paris-Saclay, F-91191 Gif-sur-Yvette, France}

\author{R.~Zhou\orcidlink{0000-0001-5381-4372}}
\affiliation{Lawrence Berkeley National Laboratory, 1 Cyclotron Road, Berkeley, CA 94720, USA}

\begin{abstract}
Joint analyses of high-resolution CMB temperature maps with galaxy surveys provide a unique way to reconstruct the radial velocity field of the underlying matter distribution via the kinematic Sunyaev–Zeldovich (kSZ) effect. Using data from the Atacama Cosmology Telescope (ACT) DR6 and the Dark Energy Spectroscopic Instrument (DESI) DR2, we present radial velocity reconstructions for luminous red galaxies (LRGs), emission-line galaxies (ELGs), and quasars (QSOs). Leveraging the spectroscopic data, we are able to reliably model the foreground contamination and report a negligible impact on our main observables. We detect the velocity-galaxy cross-correlation at $17\sigma$ for LRGs, and for the first time, at $8.3\sigma$ for ELGs and $6.8\sigma$ for QSOs. We further report the first detection of the velocity-velocity correlation using LRGs at $3.1\sigma$, as well as the highest cumulative detection of the kSZ effect to date at $20.8 \sigma$. Similarly to previous results, we find a lower amplitude of the kSZ signal compared to our fiducial halo model prediction and electron profile assuming a Battaglia profile. Combining these new observables, we obtain constraints on local-type primordial non-Gaussianity (PNG): $f_{\rm NL}^{\rm loc} = 15.9_{-34.4}^{+34.6}$ at 68\% confidence, which represents the tightest constraint to date derived from the velocity field. The measurements presented here already exhibit lower noise on a per-mode basis than the galaxy auto-correlation on the largest scales, $k \lesssim 0.002~\rm{Mpc^{-1}}$, highlighting the key role these observables will play in the context of future CMB experiments such as the Simons Observatory. 
\end{abstract}

\maketitle

\section{Introduction}
The kinetic Sunyaev-Zeldovich (kSZ) effect \cite{Sunyaev1972,Sunyaev1980,Rephaeli1991,Birkinshaw1999} provides a practical route to reconstruct the large-scale velocity field of the Universe by correlating high-resolution CMB temperature maps with galaxy surveys \cite{Deutsch2018,Smith2018,Giri2022,Cayuso2023}. Recent joint analyses of the Atacama Cosmology Telescope (ACT) CMB temperature maps \cite{Naess2025} and DESI Legacy Imaging Surveys (DESI-LS) galaxy counts \cite{Dey2019} demonstrate that this program is becoming observationally mature, where three‑dimensional \cite{Hotinli2025} and tomographic \cite{Lai2025,McCarthy2025,Villagra2026} reconstructions achieve a detection signal-to-noise of ${\sim 11.7}$ for the galaxy–velocity cross signal and prefer a low kSZ `velocity/optical‑depth' amplitude, consistent with feedback-suppressed electron profiles in massive halos and galaxy groups \cite{Hadzhiyska2025,Pandey2025,Hadzhiyska2025a}. 

Beyond detection, reconstructed velocities enable cosmological tests on ultra-large scales \cite{Munchmeyer2019,Hotinli2019,AnilKumar2022,Hotinli2022,AnilKumar2023,Hotinli2023, Adshead2024,Tishue2025,Adolff2025}, including competitive constraints on local-type primordial non‑Gaussianity (PNG) \cite{Verde2000,Maldacena2003,Bartolo2004} from the scale‑dependent response of the galaxy-velocity correlation \cite{Dalal2008,Slosar2008} as proposed recently \cite{Bloch2024,McCarthy2025a,Krywonos2024,Lague2024}.

The PNG of the local type is defined by a non-Gaussian primordial potential $\Phi$ of the form \cite{Komatsu2001}
\begin{equation}
    \Phi(\vb{x})=\phi(\vb{x}) + \fnl (\phi(\vb{x})^2-\langle\phi^2\rangle)\,,
\end{equation}
where $\phi$ is a Gaussian field and $\fnl$ quantifies deviations from Gaussian initial conditions. The most stringent bounds on $\fnl$ come from CMB bispectrum measurement using Planck data ($\fnl=0.8\pm5.0$ \cite{Akrami2020,Jung2025}) while the tightest bounds from galaxy clustering data, $\fnl = -3.6^{+9.0}_{-9.1}$ at $68\%$ confidence~\cite{Chaussidon2024a}, is from LRG and QSO of the DESI DR1 data \cite{DESIDR1} (\eg \cite{Kurita2023, Cagliari2025, Chiarenza2025, Chudaykin2025, Bermejo-Climent2026} for other recent complementary constraints). 

Thanks to cosmic variance cancellation \cite{Munchmeyer2019}, the joint analysis of the galaxy–galaxy and galaxy–velocity power spectra has the potential to provide one of the most sensitive probes of local primordial non-Gaussianity. In particular, upcoming surveys \cite{TheSimonsObservatoryCollaboration2018,LSSTScienceCollaboration2009,Dore2014} are expected to achieve constraints at the level of $\sigma(f_{\mathrm{NL}}) \sim 1$, thereby enabling robust discrimination among competing inflationary scenarios \cite{DePutter2017}. In particular, slow-roll single field inflation predicts a tiny amount of local PNG: $\fnl \sim 0.01$ \cite{Creminelli2004}.

In this work, we extend the work of \cite{Lague2024,Hotinli2025} by using the large-scale clustering of the Luminous Red Galaxy (LRG), Emission Line Galaxy (ELG) and quasar (QSO) samples from the second data release (DR2) \cite{DESIDR2} of the Dark Energy Spectroscopic Instrument (DESI) \cite{DESI2016I,DESI2016II}. At the time of writing, the galaxy-galaxy power spectrum is still blinded and cannot be included in this analysis. Once allowed, we will combine it with the result of this paper to take advantage of gains from sample variance cancellation \cite{Munchmeyer2019}. Note the two other companion analyses: \cite{Hotinli2026} explores the synergies between spectroscopic and photometric LRG samples, quantifying the impact of photometric redshift uncertainties and highlighting how well-understood redshift error distributions can enable competitive kSZ measurements, while providing guidance for upcoming surveys such as the Legacy Survey of Space and Time (LSST); and \cite{Chen2026} analyzes the DESI DR2 BGS sample and the connection between the kSZ velocity bias and the stacking technique \cite[\textit{e.g.}][]{Guachalla2025}.

This paper is organized as follows: In \cref{sec:preliminaries} we introduce the quantities measured, while the different estimators used in this analysis are introduced in \cref{sec:pipelines}. We present the data in \cref{sec:data}, and the results using our methodology in \cref{sec:results}. Finally, we conclude in \cref{sec:conclusion}.

\section{Preliminaries} \label{sec:preliminaries}
\subsection{Kinetic Sunyaev-Zeldovich Effect}
The kinematic Sunyaev-Zeldovich (kSZ) effect arises from the Doppler shift of CMB photons scattered by free electrons moving with bulk peculiar velocities. Its contribution to the CMB temperature can be written as \cite{Sunyaev1980}
\begin{equation} \label{eqn:ksz}
T_{\rm kSZ}(\boldsymbol{\theta})=\int \dd \chi K(\chi) v_r(\chi \boldsymbol{\theta}) \delta_e(\chi \boldsymbol{\theta}) \,,
\end{equation}
where $\chi$ is the comoving distance, $\boldsymbol{\theta}$ the line-of-sight, $\delta_e$ the electron density contrast, $v_r$ the radial velocity of the matter field and $K$ the kSZ radial weight function (in $\mu$K-Mpc$^{-1}$):
\begin{equation}
K(z) \equiv - T_{\rm CMB} \sigma_T n_{e,0} x_e(z) e^{-\tau(z)}(1+z)^2 \,,
\end{equation}
with $T_{\rm CMB}$ the mean CMB temperature today, $\sigma_T$ the Thomson cross-section, $n_{e,0}$ is the mean electron density today, $x_e$ is the ionization fraction and $\tau$ is the optical depth.

\subsection{Galaxies in redshift space}
At relatively mild non-linear scales ($k < 0.2~h\mathrm{Mpc}^{-1}$), the galaxy density contrast in redshift space $\delta_g^s$ is related to the linear matter contrast density $\delta_m$ in real space by taking into account the Kaiser \cite{Kaiser1987} and Fingers-of-God \cite{Percival2009} effects:
\begin{equation} \label{eqn:density_rsd}
    \delta_g^s(k, \mu) = \left(b_1 + f \mu^2\right)D_g(k, \mu, \Sigma_s)\delta_m(k) \,,
\end{equation}
with 
\begin{equation}
D_g(k, \mu, \Sigma_s) = 1 / \left(1 + (k\mu\Sigma_s)^2/2\right) 
\end{equation}
chosen to have a Lorentzian form. $b_1$ is the linear galaxy bias, $f$ the linear growth rate, $\mu \equiv \vu{k} \cdot \vu{l}$ is the directional cosine between the line-of-sight $\vu{l}$ and the wavelength $\vb{k}$, and $\Sigma_s$ quantifies the amount of damping due to the Finger-of-God effect, extending the validity of the linear description to mildly non-linear scales by absorbing the nonlinearities. When marginalized over, this damping term also effectively captures observational effects such as redshift uncertainties and fiber collisions.

\subsection{Large scale dependent bias}
In the presence of local-type primordial non-Gaussianity, long-wavelength gravitational potential fluctuations modulate small-scale structure formation, inducing a scale-dependent correction to the linear bias of biased tracers \citep{Dalal2008, Slosar2008} such that
\begin{equation} \label{eqn:scale_depedent_bias}
b_1(z) \rightarrow b_1(z) + \dfrac{b_{\Phi}(z)}{T_{\Phi \rightarrow \delta}(k, z)} \fnl \,,
\end{equation} 
where $b_{\Phi}(z)$ is the PNG bias parameter describing the response of the tracer abundance to long-wavelength potential fluctuations. $T_{\Phi \rightarrow \delta}(k, z)$ is the transfer function between the primordial gravitational field $\Phi$ and the matter density perturbation $\delta_m$:
\begin{equation}
    T_{\Phi \rightarrow \delta}(k, z) = \sqrt{\dfrac{P_{\rm lin}(k, z)}{P_{\Phi}(k)}} \,,
\end{equation}
where $P_{\rm lin}$ is the linear power spectrum of the matter at redshift $z$ and $P_{\Phi}$ the primordial potential\footnote{$\Phi$ is normalized to $3/5 \mathcal{R}$, where $\mathcal{R}$ is the comoving curvature perturbation, to match the usual definition of \cite{Slosar2008}.} power spectrum,
\begin{equation}
    P_{\Phi}(k) = \dfrac{9}{25} \dfrac{2 \pi^2}{k^3} A_s \left(\dfrac{k}{k_{\rm pivot}}\right)^{n_s-1} \,,
\end{equation}
where $n_s$ is the spectral index and $A_s$ the amplitude of the initial power spectrum at $k_{\rm pivot} = 0.05\;\text{Mpc}^{-1}$. In the following, we will fix the cosmology to Planck18 + BAO from \cite{Planck18}\footnote{Last column of Table 2.}.

While the theoretical prescription of the PNG bias $b_{\Phi}$ is a widely discussed topic \cite{Biagetti2017, Barreira2020a, Fondi2023, Sullivan2023, Adame2024, Dalal2025, Hadzhiyska2025c, Perez2026}, we will assume the simple relation
\begin{equation} \label{eqn:b_phi_with_p}
    b_{\Phi}(z) = 2 \delta_c \times (b_1(z) - p) \,,
\end{equation}
where $\delta_c = 1.686$ is the spherical collapse linear over-density and $p = 1.0$ for LRGs and ELGs (universality relation), and $p=1.4$ for the quasars \cite{Fondi2025}. We note that further study will be required for LRGs and ELGs.

\subsection{Velocities in redshift space}
On large scales, extending into the mildly non-linear regime, the radial peculiar velocity field is directly related to the matter density field through the continuity equation and therefore provides an unbiased probe of large-scale density fluctuations:
\begin{equation}
    v_r(k, \mu) = \dfrac{i f a H \mu}{k}\,\delta_m(k) \,,
\end{equation}
where $a$ is the scale factor and $H$ the Hubble expansion rate. 

Then, the redshift-space radial velocity is well approximated by \cite{Koda2014, Dam2021}
\begin{equation} \label{eqn:velocity_rsd_no_bv}
    v_r^s(k, \mu) =  \dfrac{ifaH \mu}{k}  D_{v}(k, \Sigma_v) \delta_m(k)
\end{equation}
where 
\begin{equation}
    D_v(k, \Sigma_v) = \mathrm{sinc}(k\Sigma_v)
\end{equation}
is a damping function controlled by $\Sigma_v$ to account for the redshift space distortion. 

Finally, as discussed in \cref{sec:velocity_bias}, the reconstructed radial velocity field is, in general, a biased estimate of the true peculiar velocity due to uncertainties in the small-scale distribution of ionized gas that sources the kSZ signal. We therefore introduce a scale-independent amplitude parameter in \cref{eqn:velocity_rsd_no_bv},
\begin{equation} \label{eqn:velocity_rsd}
    v_r^s(k, \mu) \rightarrow b_v\, v_r^s(k, \mu) \,,
\end{equation}
where $b_v$ denotes the effective \emph{kSZ velocity bias} (often interpreted as an optical-depth bias), and is treated as a nuisance parameter marginalized over in the inference. If the response of the kSZ signal, \ie the galaxy-electron power spectrum, is perfectly known, $b_v = 1$, see \cref{sec:velocity_bias}.

\subsection{Multipoles of the Power spectrum}
In the following, we will work with the Legendre multipoles of the power spectrum $P^{ab}(\vb{k}) = \langle a(\vb{k}) b(\vb{k)} \rangle$:
\begin{equation}
    P_{\ell}^{ab}(k) = \dfrac{2\ell + 1}{2} \int_{-1}^{1} \dd\mu P^{ab}(k,\mu) \mathcal{L}_{\ell}(\mu) \,,
\end{equation}
where $a$ and $b$ will be either the galaxy contrast $\delta_g^{s}$ from \cref{eqn:density_rsd} or the radial velocity $v_{r}^{s}$ from \cref{eqn:velocity_rsd}.

Notably, neglecting the damping term ($\Sigma_g=0, \Sigma_v=0$), the only non-zero multipoles for the galaxy-velocity power spectrum are the dipole ($\ell=1$) and the octopole ($\ell=3$)\footnote{The galaxy damping term is even and will only create a tiny non-zero pentadecapole ($\ell=5$) multipole but no even multipoles.}:
\begin{align} \label{eqn:pgv_theo}
    P_{\ell=1}^{gv} &= \left(b_1 + \frac{3}{5}f \right) b_v \frac{ifaH}{k} P_{\rm lin}(k) \,, \\
    P_{\ell=3}^{gv} &= \frac{2}{5} b_v \frac{if^2aH}{k} P_{\rm lin}(k) \,.
\end{align}

Similar to the well-known case of the galaxy-galaxy power spectrum, the velocity-velocity power spectrum will only have non-zero even multipoles. In particular, the monopole reads as
\begin{align} \label{eqn:pvv_theo}
    P_{\ell=0}^{vv} &= - \dfrac{1}{5}b_v^2\left(\dfrac{faH}{k}\right)^2 P_{\rm lin}(k)\,.
\end{align}

Note that $\delta_m^s$ and $v_r^s$ are redshift dependent. As usual, we will work with an effective redshift $z_{\rm eff}$\footnote{$z_{\rm eff} = \left(\int \dd z n(z)^2 w_a(z)w_b(z) z\right) / \left( \int \dd z  n(z)^2 w_a(z)w_b(z)\right)$ where $w_a$ and $w_b$ are the weights of two fields that are cross-correlated.} (see \cref{tab:desi_tracers}), such that the different parameters describing the tracers will depict their effective quantities. 

Finally, the observed power spectra are generally described using the window matrix formalism \cite{Beutler2021} to account for the survey geometry, masks, and the integral constraint \cite{DeMattia2019}, and this approach is typically adopted when measuring the largest-scale modes of the galaxy power spectrum \cite{Chaussidon2024a}. Here, we do not follow this approach and instead rely on a surrogate-field methodology, introduced in \cite{Hotinli2025}, in which model predictions are forward-modeled through the survey geometry and masks, as described in \cref{sec:surrogates}.
\section{Description of our pipeline} \label{sec:pipelines}
The pipeline used for this analysis is based on the previous works \cite{Smith2018,Hotinli2025} and was updated on several points. In particular, we now include redshift space distortion, foregrounds contamination and allow high-order Legendre multipoles computation. Our pipeline, \texttt{kszx}, is publicly available\footnote{\url{https://github.com/echaussidon/kszx}}. 

\subsection{kSZ Velocity Reconstruction Estimator} \label{sec:ksz_reconstruction}
We reconstruct the large-scale radial velocity field using the quadratic kSZ estimator introduced in \cite{Smith2018} and implemented in \cite{Hotinli2025}. The inputs are the three-dimensional galaxy overdensity $\delta_g(\mathbf{x})$ and the two-dimensional CMB temperature map $T_{\rm CMB}(\boldsymbol{\theta})$, and the output is a three-dimensional field $\hat v_r(\mathbf{x})$ which traces the true radial velocity on large scales up to an overall normalization.

In a simplified flat-sky ``snapshot'' geometry with no mask, the estimator can be written schematically as
\begin{equation}
    \hat v_r(\mathbf{k}_L) \propto
    \int_{\mathbf{k}_S + \mathbf{l}/\chi_* = \mathbf{k}_L}
    \frac{P_{ge}(k_S)}{P_{gg}^{\rm tot}(k_S)} 
    \frac{1}{C_\ell^{\rm tot}}
    \, \delta_g(\mathbf{k}_S)\, T_{\rm CMB}(\mathbf{l})
\end{equation}
where $\chi_*$ is the comoving distance to the galaxy sample, $P_{ge}$ is the galaxy–electron power spectrum, $P_{gg}^{\rm tot}$ is the total galaxy power spectrum including shot noise, and $C_\ell^{\rm tot}$ is the total CMB power spectrum including foregrounds and noise. The integral is dominated by squeezed configurations $k_L \ll k_S$ where $L$ (\textit{resp.} $S$) denotes long (\textit{resp.} short) modes. $P_{ge}$ is computed with the halo model prescription in \texttt{hmvec}\footnote{\url{https://github.com/simonsobs/hmvec}} using the Battaglia gas profile \cite{Battaglia2016}. See \cref{sec:low_bv} for a discussion about this choice.

\begin{figure}
    \centering
    \hspace{-3mm}\includegraphics[scale=1]{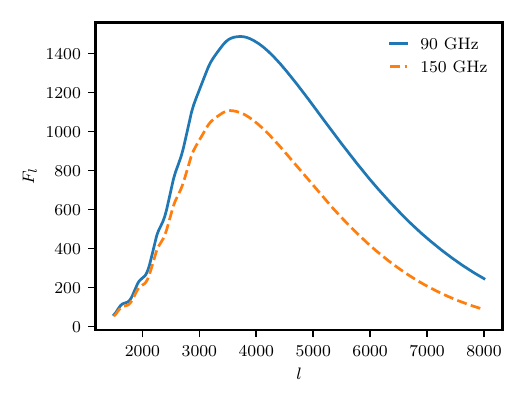}
    \caption{CMB filters $F_{l}$ (\cref{eqn:cmb_filter}) used for the LRG case. The two filters peak around $l \sim 4000$ where the kSZ signal is the strongest and where our signal is coming from. First modes ($\ell \leq 1500$) are filtered out because they contain only the primary CMB information, while smaller scales contain only noise.}
    \label{fig:filters}
\end{figure}

Practically, this estimator is implemented in real space:
\begin{equation} \label{eqn:velocity_estimator}
\widehat{v}_r(\vb{x})=\sum_b \sum_{i \in b} W_i^v\left(\widetilde{T}\left(\boldsymbol{\theta}_i\right)-\widetilde{T}_b\right) \delta^3\left(\vb{x}-\vb{x}_i\right)
\end{equation}
where we sum over multiple redshift bins $b$, $W_i^v$ is a per-galaxy weight (see \cref{sec:data}) and $\widetilde{T}$ is the all-sky filtered CMB temperature whose spherical harmonics\footnote{$\widetilde{T}(\boldsymbol{\theta}) =\sum_{l m} \widetilde{T}_{l m} Y_{l m}(\boldsymbol{\theta})$.} are
\begin{equation} \label{eqn:filter_temperature}
  \widetilde{T}_{l m} =F_l \int \dd^2 \boldsymbol{\theta} W_{\rm CMB}(\boldsymbol{\theta}) T_{\mathrm{CMB}}(\boldsymbol{\theta}) Y_{l m}^*(\boldsymbol{\theta})
\end{equation}
with $Y_{l m}$ the spherical harmonic functions, $W_{\rm CMB}$ is a CMB pixel weight (see \cref{sec:data}) used to downweight regions with high noise or large foreground contribution, and $F_l$ the actual filter defined as
\begin{equation} \label{eqn:cmb_filter}
 F_l \equiv \frac{P_{ge}^{\rm fid}\left(l / \chi_*\right)}{P_{gg}^{\rm fid}\left(l / \chi_*\right)} \frac{b_l}{C_l^{\mathrm{tot}}}
\end{equation}
where we also include in the filter the CMB beam $b_{l}$. The filters for the LRG case are displayed in \cref{fig:filters}. 

To reduce the CMB foregrounds as empirically found in \cite{Hotinli2025}, we subtract in \cref{eqn:velocity_estimator} the mean filtered CMB temperature in 25 redshift bins, where the mean is $\widetilde{T}_b = \sum_{i\in b} W_i^{v} \widetilde{T}(\boldsymbol{\theta}_i) / \sum_{i\in b} W_i^{v}$. However, this subtraction removes only additive contributions that are slowly varying in redshift and correlated with the galaxy sample. It does not fully eliminate foreground-induced correlations, which are modeled explicitly in \cref{sec:foreground_model}.

This procedure is analogous to the radial integral constraint \cite{DeMattia2019}, where the mean radial density is fixed within each redshift bin. As in that case, the subtraction modifies the estimator and induces a bias in the measured power spectrum. In the following, this effect is consistently modeled using surrogate realizations (see \cref{sec:surrogates}).

\subsection{kSZ Velocity Bias} \label{sec:velocity_bias}
As shown in Appendix D of \cite{Hotinli2025}, \cref{eqn:velocity_estimator} is a biased estimate of the true radial velocity which satisfies
\begin{equation} \label{eqn:estimated_v}
\left\langle\widehat{v}_r(\vb{x})\right\rangle=b_v B(\vb{x}) n_v(\vb{x}) v_r^{\rm true}(\vb{x}) \,,
\end{equation}
where $n_v$ is the 3d galaxy density weighted with the per-object velocity weight $W_v$, and
\begin{equation} \label{eqn:B}
B(\chi, \boldsymbol{\theta}) = W_{\rm CMB}(\boldsymbol{\theta}) \frac{K(\chi)}{\chi^2} \int \frac{\dd^2 \vb{L}}{(2 \pi)^2} b_L F_L P_{g e}^{\mathrm{fid}}(L / \chi) \,.
\end{equation}

In the following, $W_{\rm CMB}$ is chosen to be 1 in the common area between CMB temperature and DESI galaxies, see \cref{sec:act_data} and we will compute this quantity at $z_{\rm eff}$ leading to a constant quantity: 
\begin{equation} \label{eqn:B_value}
B \simeq -6.43\times10^5, -4.49\times10^4, -3.41\times10^5~\mu\rm{K}^{-1}
\end{equation}
for LRG, ELG and QSO. It is worth noting that B is negative by construction, given the convention adopted in \cref{eqn:ksz}.

Finally, $b_v$ is known as the \emph{kSZ velocity bias} that account for mismatch between the fiducial and true small-scale galaxy-electron cross power spectra, and has the form:
\begin{equation} \label{eqn:velocity_bias}
b_v(\chi)=\frac{\int \dd^2\mathbf{L}~ b_L F_L P_{g e}^{\text {true }}(L/\chi)}{\int \dd^2\mathbf{L}~ b_L F_L P_{g e}^{\text {fid }}(L/\chi)}\,.
\end{equation} 
As shown in Ref.~\cite{Giri2022}, on the scales relevant for our analysis $b_v$ is effectively constant\footnote{Formally, $b_v$ is constant up to $0.04 ~\rm{Mpc}^{-1}$ (see Fig.3 of \cite{Giri2022}), the residual scale dependence on smaller scales will be naturally absorbed with the damping term included in \cref{eqn:velocity_rsd_no_bv}.}, such that it rescales the overall amplitude of the reconstructed kSZ signal, as introduced in \cref{eqn:velocity_rsd}. We therefore treat $b_v$ as a free amplitude parameter and marginalize over it in the inference. In the case where the galaxy-electron power spectrum is known \ie $P_{ge}^{\rm fig} = P_{ge}^{\rm true}$, $b_v=1$.

We note that the estimated radial velocity, \cref{eqn:velocity_estimator}, is in Mpc$^{-3}$ $\mu$K$^{-1}$ units whereas the physical radial velocity field $v_r$ is dimensionless (in units of $c$). In the following, these normalization and unit conventions are consistently propagated into our surrogate methodology, see \cref{sec:surrogates}, ensuring that the measured and predicted quantities are defined and compared on an identical footing.

\subsection{Foreground Contributions} \label{sec:foreground_model}
In the previous sections, we implicitly assumed that the filtered CMB temperature $\widetilde{T}$ of \cref{eqn:filter_temperature} carries only kSZ information. This is, of course, not strictly true because, on these scales, the primary CMB anisotropies, secondary anisotropies such as the thermal SZ effect \cite{Zeldovich1969}, and other foregrounds such as the cosmic infrared background (CIB) \cite{Hauser2001} also contribute significantly:
\begin{equation}
    T_{\rm obs} = T_{\rm prim} + T_{\rm kSZ} + T_{\rm tSZ} +  T _{\rm{CIB}}\,.
\end{equation}

The estimator in \cref{eqn:velocity_estimator} combines small-scale galaxy fluctuations with the observed filtered CMB temperature $\tilde{T}_{\rm obs}$ evaluated at galaxy positions \ie it is sensitive to the galaxy-temperature correlation $\langle \delta_g\,\widetilde{T}_{\rm obs}\rangle$ on the angular scales selected by the filter. On these scales, the primary CMB anisotropies are statistically uncorrelated with the galaxy density and therefore contribute only to the reconstruction noise. The remaining correlated contributions arise from the kSZ signal (see \cref{sec:velocity_bias}) and from foregrounds that trace large-scale structure, most notably the combination of tSZ and CIB. 

Both of the tSZ and CIB contributions can be viewed as a biased tracer of the matter density field, $b_{\rm fg}\delta_m$, where $b_{\rm fg}$ quantifies the amplitude of this foreground contribution (see \cref{app:foreground_model} for a halo-model derivation). This will be validated with great success in \cref{sec:lrg_foregrounds}. Such that our estimator reads as 
\begin{equation} \label{eqn:velocity_rsd_fg}
\widehat{v}_r(\vb{x})=b_{\rm fg}\delta_{m}(\vb{x}) + b_v B(\vb{x}) n_v(\vb{x}) v_r^{\text {true }}(\vb{x}) + \rm{noise} \,,
\end{equation}
where we have introduced a noise term that does not bias the estimator but increases the variance of the measurements, and will be modeled in \cref{sec:surrogates}. 

Hence, in the galaxy-velocity power spectrum, the two terms $\delta_m$ and $v_r^{\rm true} \propto \mu \delta_m$ will be naturally projected onto different multipoles due to their distinct $\mu$-dependence: the density term contributes to even multipoles, while the velocity term contributes to odd multipoles. Without any survey geometry, these two contributions can be separated such that foregrounds do not bias the kSZ signal. However, the survey selection function (finite size of the survey, angular masks, completeness, radial selection, etc. leads to the usual window function (\eg \cite{Beutler2021}) with which the theory must be convolved. This convolution mixes multipoles, allowing foreground power from even multipoles (\eg the monopole) to leak into odd multipoles (\eg the dipole), thereby contaminating the velocity signal.

In the following, we model this contamination and multipole leakage using the surrogate methodology described in \cref{sec:surrogates}, and infer the foreground amplitude $b_{\rm fg}$ from the monopole of the galaxy-velocity power spectrum (see \cref{sec:lrg_foregrounds}).

Furthermore, we neglect potential biases from other secondary CMB anisotropies such as gravitational lensing and ISW-type contributions. Gravitational lensing acts as a remapping of the primary CMB and is odd under $T_{\rm prim}\rightarrow -T_{\rm prim}$ transformation; since our estimator is linear in the filtered temperature, this symmetry nulls any lensing bias. Linear ISW arises on very large angular scales ($\ell < 100$) and is strongly suppressed by our filtering. It will add some positive contribution to the monopole, similarly to the CIB, but only at very large scales. The nonlinear ISW (Rees--Sciama effect) is intrinsically much smaller than the kSZ signal, while the moving-lens effect is also small, and depends on transverse velocities, hence is odd under reversal of the transverse velocity. We therefore neglect these contributions as sources of bias and treat them only as an additional source of reconstruction noise.

\subsection{Power Spectrum Estimator} \label{sec:pk_estimator}
First let's define the spin-$\ell$ Fourier transform of a field $f$ as\footnote{Conventions of our FFT are defined here: \url{https://kszx.readthedocs.io/en/latest/fft.html\#ffts-with-spin}}
\begin{equation} \begin{cases} \label{eqn:spin-l}
\mathcal{F}_{\ell}\left[f\right](\vb{k}) &= \displaystyle\int \dd \vb{x} \epsilon^{*} \mathcal{L}_{\ell}(\mu) f(\vb{x})e^{-i\vb{k}\cdot\vb{x}} \vspace{2mm}  \\ 
\mathcal{F}_{\ell}\left[f\right](\vb{x}) &= \displaystyle\int \dfrac{\dd \vb{k}}{(2\pi)^3} \epsilon \mathcal{L}_{\ell}(\mu) f(\vb{k}) e^{i\vb{k}\cdot\vb{x}}
\end{cases} \end{equation}
where 
\begin{equation}
\epsilon= \begin{cases}i & \text { if } l \text { is odd } \\ 1 & \text { if } l \text { is even }\end{cases}.
\end{equation}
The chosen definition of \cref{eqn:spin-l} is practical since it relates $\delta_m$ and $v_r$ via 
\begin{equation}
v_r(\vb{x})= \mathcal{F}_{\ell=1}\left[\vb{k} \mapsto \frac{f a H}{k}\delta_m(\vb{k})\right].
\end{equation}

To speed up its computation, this transformation is decomposed into $(2\ell+1)$ FFTs as in the case of the usual power spectrum estimators \cite{Hand2017}. However, our definition is slightly different than Eq.~(2.7) of \cite{Hand2017}. 

Built on this spin-$\ell$ Fourier transform, the galaxy-velocity power spectrum estimator \cite{Yamamoto2006, Hand2017} reads as
\begin{equation} \label{eqn:pgv_estimator}
     \widehat{P}_{\ell}^{gv}(k) = \dfrac{2\ell + 1}{\mathcal{N}_{gv}} \int \dfrac{\dd \Omega_k}{4\pi} \mathcal{F}_{0}\left[\widehat{\delta}_g\right](\vb k) \mathcal{F}_{\ell}\left[\widehat{v}_r\right](-\vb k)\,,
\end{equation}
where $\widehat{\delta}_g$ is the Feldman-Kaiser-Peacock (FKP) galaxy density field \cite{Feldman1994}, $\widehat{v}_r$ is given in \cref{eqn:velocity_estimator}, and $\mathcal{N}_{gv}$ is a normalization factor. It is chosen to follow the standard FKP convention as explained in Appendix A of \cite{Hotinli2025} and it is estimated from the randoms\footnote{See \url{https://kszx.readthedocs.io/en/latest/wfunc_utils.html\#details-of-w-crude}, where the purpose of the subtraction is to cancel the shot noise from the randoms.}. We include $B(\vb{x})$ in the normalization ensuring the galaxy-velocity power spectrum is expressed in $\rm{Mpc}^3$ (in units of $c$) and does not depend on the sign convention in \cref{eqn:ksz}.

The definition of \cref{eqn:spin-l}, the estimated $\widehat{P}_{\ell}^{gv}$ is real \textit{i.e.} does not have the $i$ as in \cref{eqn:pgv_theo}.  In our case, this choice or the choice of the normalization factor do not really matter since the theory will be calculated using the surrogate methodology, see \cref{sec:surrogates}, that will take it into account consistently. 

Formally, \cref{eqn:estimated_v} provides an estimation of the momentum radial velocity $p(\vb{x}) = \left(1 + \delta_g(\vb{x}) \right) v_r(\vb{x})$ and therefore \cref{eqn:pgv_estimator} corresponds to the galaxy--momentum velocity power spectrum $P_{\ell}^{gp}(k)$ \cite{Howlett2019}. This arises because the velocity field is sampled at the discrete positions of galaxies. On linear scales --which are precisely the scales probed by the kSZ velocity reconstruction--, the two power spectra are equivalent: $P_{\ell}^{gp}(k)$ = $P_{\ell}^{gv}(k)$ \cite{Park2000}.

Similarly, the estimator of the velocity-velocity power spectrum reads as
\begin{equation} \label{eqn:pvv_estimator}
\widehat{P}_{\ell,\ell^{\prime}}^{vv}(k) = \dfrac{(2\ell+ 1)(2\ell^{\prime} + 1)}{\mathcal{N}_{vv}} \int \dfrac{\dd \Omega_k}{4\pi} \mathcal{F}_{\ell}\left[\widehat{v}_r\right](\vb{k}) \mathcal{F}_{\ell^{\prime}}\left[\widehat{v}_r\right](-\vb k)\,.
\end{equation}
With this notation, the usual monopole of the velocity-velocity power spectrum (\cref{eqn:pvv_theo}) is $-\widehat{P}_{\ell=0, \ell=0}^{vv}$. However, in the following, we prefer to use $\widehat{P}_{\ell=1,\ell=1}^{vv}$ since it is the optimal way to weight the $\mu$-dependence in the velocity term \cite{Tegmark1997}. An additional advantage is the reduced impact of the foreground contribution in this estimator. A comparison between the two choices is presented in \cref{app:ell00vsell11}.

In the following, the galaxy-velocity and velocity-velocity power spectrum estimators are computed with the binning specified in \cref{tab:binning}. At small scales, wider bins are adopted to reduce cross-covariance and the total number of data points, while at large scales thinner bins are used to enhance sensitivity to $\fnl$.

With these choices and using the two CMB temperature maps described below (90 and 150\,GHz), our data vector size will be 56 for the galaxy-velocity power spectra (g$\times$90 and g$\times$150) and 63 for the velocity-velocity power spectra (90$\times$90, 90$\times$150 and 150$\times$150).

\begin{table}[t!]
    \centering
    \caption{Binning used to compute the estimator of the galaxy-velocity and the velocity-velocity power spectra.}
    \vspace{1.5mm}
    \begin{tabular}{lcc}
    \toprule
    $\left[\text{Mpc}^{-1}\right]$ & $\widehat{P}_{\ell}^{gv}$ & $\widehat{P}_{\ell,\ell^\prime}^{vv}$ \\
    \midrule
       $k_{\rm min}$ & 0.0014 & 0.0014 \\
       $k_{\rm max}$ & 0.07 & 0.05 \\
       $\Delta k \text{ (for }k < 0.01)$ & 0.0014 & 0.0014\\
       $\Delta k \text{ (for }k \geq 0.01)$ & 0.0028 & 0.0028\\
       $N_{k}$  & 28 & 21 \\
    \bottomrule
    \end{tabular}
    \label{tab:binning}
\end{table}

\subsection{Surrogate Fields} \label{sec:surrogates}
\subsubsection{Model and covariance predictions}
To model the observed power spectrum and the associated covariance matrix, we rely on the surrogate methodology. We refer the reader to \cite{Hotinli2025}, and in particular to its appendix for a mathematical justification. The main advantage of this methodology is the inexpensive generation of the covariance matrix with parameter dependence, which consistently incorporates survey geometry and window effects without requiring large ensembles of realistic simulations. Our estimated covariance is a good approximation at large scales ($k < 0.07$ Mpc$^{-1}$) where the higher-order contributions are negligible.

A surrogate field is a fast Gaussian field painted onto the randoms associated with the data with the two-point statistical properties. At linear scales, a set of surrogates therefore reproduces the correct covariance while sharing the same survey geometry as the data. Note that a surrogate does not resemble a mock catalog: rather than simulating a clustered catalog with realistic higher-order statistics, it provides a simplified field-level realization with the correct covariance, obtained by ``painting'' a Gaussian field onto the randoms (and adding a noise term). The power spectra computed from the surrogates can be used to estimate the covariance matrix, as well as to predict the observed power spectrum from the data. 

A surrogate field mimics the galaxy density field \cref{eqn:density_rsd} and the velocity field \cref{eqn:velocity_rsd_fg} as\footnote{This simple expression is valid because we consider only linear contributions. More generally, this corresponds to performing a Taylor expansion around the fiducial cosmology.} 
\begin{equation} \label{eqn:surrogate}
\begin{cases}
&S_{\ell}^{g}(\vb{k}) = \mathcal{F}_{\ell}\left[\nabla S_g\right] \cdot p_g \\[1ex]
&S_{\ell}^{\rm freq}(\vb{k}) = \mathcal{F}_{\ell}\left[\nabla S_{\rm freq} \right] \cdot p_{\rm freq}
\end{cases},
\end{equation}
where 
\begin{equation}
\begin{cases}
&p_g = \left(b_1D_g(k), fD_g(k), b_{\Phi}\fnl D_g(k), s_n \right) \\[1ex]
&p_{\rm freq} = \left(b_{\rm fg}^{\rm freq}, b_v^{\rm freq}D_v(k), s_{nv}^{\rm freq} \right)
\end{cases}\,.
\end{equation}
The quantities $\nabla S_g$, $\nabla S_{\rm freq}$ are explained in \cref{sec:surrogate_creation}.

For each surrogate realization, the power spectrum of two fields $f_1, f_2$ is evaluated using either \cref{eqn:pgv_estimator} or \cref{eqn:pvv_estimator}. The predicted observable power spectrum is then the mean over the surrogate realizations. The covariance matrix is estimated from the numerical covariance of this ensemble. To accelerate the inference, we precompute the basis contributions to both the model and the covariance (\eg each term from $\mathcal{F}_{\ell}\left[\nabla S_g\right] \times \mathcal{F}_{\ell}\left[\nabla S_{\rm freq} \right]$ for the galaxy-velocity power spectrum). For the observables that depend linearly on the parameter set; evaluating the prediction for a given set of parameters then reduces to simple linear combinations of these precomputed terms. 

As with simulation-based covariance estimates, we correct for the finite number of realizations by applying the Hartlap factor \cite{Hartlap2007} and the Percival factor \cite{Percival2014}. These corrections account for the bias in the inverse covariance matrix and the associated propagation of covariance uncertainty into parameter constraints arising from the inverse Wishart distribution. The latter is not included to assess the goodness of fit \ie when computing the $\chi^2$ or the PTE. 

To avoid large correction factors, we generate a sufficient number of surrogate realizations (see \cref{tab:nsurr}) such that the combined Hartlap and Percival corrections amount to approximately 5\%. Since the data vector is larger when including the velocity-velocity power spectra, we generate a correspondingly larger number of surrogates for the LRG sample.

\begin{table}[t!]
    \centering
    \caption{Number of surrogates used for the different tracers.}
    \label{tab:nsurr}
    \vspace{1.5mm}
    \begin{tabular}{cc}
    \toprule
       Tracer & $\#$ surrogates \\ \midrule
        LRG   & $4500$ \\
        ELG   & $2000$ \\
        QSO   & $2000$ \\
    \bottomrule
    \end{tabular}
\end{table}

\subsubsection{Few remarks} \label{sec:remarks}
First, we have different velocity parameters and components in \cref{eqn:surrogate} for each CMB frequency temperature map used for the velocity reconstruction because the weights, filters, and the masks involved in this reconstruction are not identical across frequencies. We therefore fit independent $b_v$ parameters for each frequency combination, while using a common $\Sigma_v$ for each tracer, as $\Sigma_v$ does not depend on the choice of the CMB map.

Second, we add the damping terms $D_g$ and $D_v$ after the spin-$\ell$ Fourier transform \ie we add them after the window convolution, directly in the multipoles. This enables us to compute the spin-$\ell$ Fourier transform of each component only once and then vary and marginalize over the damping amplitudes efficiently. We checked that the damping functions used here are able to recover the window-convolved damping behavior. The only drawback is that the parameters $\Sigma_g$ or $\Sigma_v$ are not the same in the different multipoles. This is not an issue here, since we marginalize over these parameters and do not use different multipoles to constrain them. 

In this analysis, we do not include the galaxy-galaxy power spectrum so the galaxy-velocity power spectrum alone exhibits degeneracies among $b_1$, $f$, and $b_v$, as well as between $\Sigma_g$ and $\Sigma_v$. We therefore fix $f$ to its fiducial value and fix $b_1$ using the amplitude of the galaxy-galaxy power spectrum measured separately. We fix $D_g(k) = 1$ (equivalently $\Sigma_g=0$) so that the damping in the galaxy-velocity power spectrum is controlled solely by $\Sigma_v$, where we check that the shape of the damping due to $D_g \times D_v$ can be recovered by $D_v$ alone. 

This choice will prevent us to use the same damping parameter in the galaxy-velocity and velocity-velocity power spectrum. We therefore introduce an independent parameter $\Sigma_v^{vv}$ to control the damping in the velocity-velocity power spectrum. We note that this is only a convenient choice that will not be useful once the galaxy-galaxy power spectrum is included. These damping parameters do not contain relevant information for this analysis and are  marginalized over in the inference.

Finally, we allow for a rescaling of the shot noise ($s_n$) and the velocity reconstruction noise ($s_{nv}^{\rm freq}$) to match the small-scale behavior of the measured galaxy-galaxy and velocity-velocity power spectra. We fit for three independent reconstruction-noise amplitudes ($s_{nv}^{90x90}$, $s_{nv}^{90x150}$, $s_{nv}^{150x150}$). In particular, the value $s_{nv}^{90x150}$ is used to model the cross-covariance between the two galaxy-velocity power spectra (90 and 150 GHz). For simplicity, we neglect corresponding rescaling to the cross-covariance between the 90$\times$90 and 150$\times$150 velocity-velocity power spectra.

\subsubsection{Generating one surrogate} \label{sec:surrogate_creation}
For a given surrogate, we start by generating a Gaussian field $\delta_G$ in Fourier space with the same power spectrum as the linear matter power spectrum at $z=0$. The Gaussian field is then rescaled by the growth factor to include the redshift dependence.

The different contributions to the surrogate density field can be written as
\begin{widetext}
\begin{equation} \label{eqn:surrogate_density}
\left\{
\begin{aligned}
    \nabla S_g^{b_1}(\vb{x}) &\equiv \frac{N_g}{N_r} \sum_{j \in \text{rand}} W_j^{\delta} D(z_j)\mathcal{F}_0\left[\delta_G\right]\left(\vb{x}_j\right) \delta^3\left(\vb{x}-\vb{x}_j\right), \\
    \nabla S_g^{f}(\vb{x}) &\equiv  \frac{N_g}{N_r} \sum_{j \in \text{rand}} W_j^{\delta} D(z_j)\left[\dfrac{2}{3}\mathcal{F}_{2} + \dfrac{1}{3} \mathcal{F}_0\right] \left[\delta_G\right]\left(\vb{x}_j\right)\delta^3\left(\vb{x}-\vb{x}_j\right) \\
    \nabla S_g^{\fnl}(\vb{x}) &\equiv \frac{N_g}{N_r} \sum_{j \in \text{rand}} W_j^{\delta} \mathcal{F}_0 \left[\dfrac{\delta_G}{T_{\Phi \rightarrow \delta}(k, z=0)}\right]\left(\vb{x}_j\right)\delta^3\left(\vb{x}-\vb{x}_j\right) \\
    \nabla S_g^{\mathrm{noise}}(\vb{x}) &\equiv \frac{N_g}{N_r} \sum_{j \in \mathrm{rand}} W_j^\delta \eta_j \delta^3\left(\vb{x}-\vb{x}_j\right) \text{ with } \left\langle\eta_j^2\right\rangle=\frac{N_r}{N_g}- D(z_j)^2b_1^2\left\langle\delta_G^2\right\rangle
\end{aligned}
\right. ,
\end{equation}
\end{widetext}
where the summation is performed over the randoms. $N_g$ (\textit{resp.} $N_r$) denotes the total number of galaxies (\textit{resp.} randoms), $W_j^{\delta}$ is the weight used to build the FKP galaxy density field and $D(z)$ is the growth factor normalized to unity at $z=0$. The Gaussian white noise is constructed using $\eta_j$, a Gaussian random variable, independent at each position, with variance $\left\langle \eta_j^2\right\rangle = N_r/N_g - D(z)^2 b_1^2\left \langle\delta_G^2 \right\rangle$ where $b_1$ is fixed to its fiducial value at the effective redshift of the tracer.

Similarly, the different contributions to the surrogate velocity field are
\begin{widetext}
\begin{equation} \label{eqn:surrogate_velocity}
\left\{
\begin{aligned}
\nabla S_{\rm freq}^{b_{\rm fg}}(\vb{x}) &\equiv \frac{N_g}{N_r} \sum_{j \in \mathrm{rand}} W_j^{\rm freq} B\left(\vb{x}_j\right)D(z_j) \mathcal{F}_0\left[\delta_G\right]\left(\vb{x}_j\right) \delta^3\left(\vb{x}-\vb{x}_j\right) \\
\nabla S_{\rm freq}^{b_v}(\vb{x}) &\equiv \frac{N_g}{N_r} \sum_{j \in \mathrm{rand}} W_j^{\rm freq} B\left(\vb{x}_j\right) D(z_j) faH \mathcal{F}_1\left[\dfrac{\delta_G}{k}\right]\left(\vb{x}_j\right) \delta^3\left(\vb{x}-\vb{x}_j\right) \\
\nabla S_{\rm freq}^{\rm noise}(\vb{x}) &\equiv \sum_{j \in \mathrm{rand}} M_j W_j^{\rm freq} \widetilde{T}\left(\boldsymbol{\theta}_j\right) \delta^3\left(\vb{x}-\vb{x}_j\right)
\end{aligned}
\right. ,
\end{equation}
\end{widetext}
where $W_j^{\rm freq}$ is the weighting scheme for the velocity field, $B$ is given in \cref{eqn:B} and is constant on the randoms footprint \cref{eqn:B_value}. Note that $B$ was included in $\nabla S_{\rm freq}^{b_{\rm fg}}(\vb{x})$ while it should not (see \cref{eqn:velocity_rsd_fg}), and will be removed in future analyses. Since $B$ is constant over the randoms footprint, the only consequence is the true foreground bias $b_{\rm fg}^{\rm true}$ is related to the measured $b_{\rm fg}$ through
\begin{equation} \label{eqn:true_bfg}
    b_{\rm fg}^{\rm true} \equiv B\times b_{\rm fg} \,,
\end{equation} 
where the value of B are given in \cref{eqn:B_value}.

The noise is estimated using a bootstrap strategy based on the real high-pass-filtered ACT maps $\widetilde{T}$. For each surrogate, we build a subset S of $N_g$ points randomly chosen from $\{1, \dots, N_r\}$ and define
\begin{equation}
  M_j \equiv \begin{cases}1 & \text { if } j \in S \\ 0 & \text { if } j \notin S\end{cases}\,.
\end{equation}
We use the same subset $S$ for the different frequencies to capture potential common noise.

Since the field is painted onto the randoms, the estimated power spectrum from a surrogate naturally accounts for geometrical effects such as masks (window function), wide-angle effects, and the global integral constraint \cite{Beutler2021}. However, because the redshifts of the randoms are drawn from the data (shuffling method) \cite{Ross2024}, we need to take into account the radial mode removal known as radial integral constraint \cite{DeMattia2019} \footnote{Here, we do not account for the angular integral constraint \cite{Chaussidon2024a} in this present analysis because we are not considering the large scales of the galaxy-galaxy power spectrum.} Hence, we apply a mean subtraction in 25 redshift bins to each density-field contribution defined in \cref{eqn:surrogate_density}.

Similarly, to mitigate CMB foregrounds we apply a mean-subtraction in \cref{eqn:velocity_estimator}, which introduces an additional radial integral constraint. To remain consistent and capture the impact of this mode removal, we subtract the mean in 25 redshift bins from each velocity-field contribution defined in \cref{eqn:surrogate_velocity}.

Note that using the same realization $\delta_G$, we generate the surrogate fields for the 90 and 150 GHz CMB maps. The signal components $\nabla S_{\rm freq}^{b_{fg}}(\vb{x})$ and $\nabla S_{\rm freq}^{b_v}(\vb{x})$ are therefore fully correlated between frequencies. The noise $\nabla S_{\rm freq}^{\rm noise}(\vb{x})$ is also correlated between the 90 and 150 GHz components, since $\widetilde{T}_{90}(\theta)$ and $\widetilde{T}_{150}(\theta)$ share primary CMB anisotropies and foregrounds.

\subsubsection{Computational Cost}
The surrogate methodology has a modest computational cost. As usual, the complexity and the cost of FFT scale with the mesh size. In our case, we fix the cell size to $10~\rm{Mpc}$ leading to a mesh size of 768 (\textit{resp.} 1536) for LRG (\textit{resp.} QSO) such that a single surrogate takes 210s (\textit{resp.} 250s) using 25 (\textit{resp.} 64) threads and we have sufficient memory to run 10 (\textit{resp.} 4) in parallel in a single CPU node at NERSC. We note that such a resolution in the mesh is not needed for the study of the large scales, and the cell size will be increased in further work to reduce the computational time.

With our current implementation, it is faster to run as many surrogates as possible in parallel. The computation could be significantly accelerated by using GPU to perform the FFTs, making this analysis scalable for any upcoming Stage-4 galaxy survey without methodological changes. However, we emphasize that faster execution on GPUs does not necessarily mean improved efficiency in terms of natural resource usage. For reference, the total analysis presented here (excluding preliminary runs and debugging) cost $\sim 165$ CPU hours at NERSC\footnote{At NERSC, 1,000 CPU hours are estimated to correspond to 0.38 tCO2eq.}.
\section{Data} \label{sec:data}
\subsection{ACT DR6} \label{sec:act_data}
The Atacama Cosmology Telescope (ACT) is a six-meter millimeter-wave telescope located in the Atacama Desert in Chile, designed to observe the cosmic microwave background with high angular resolution and sensitivity. We use the public ACT DR6 \cite{Naess2025} co-added source-free \texttt{daynight} temperature maps at central frequencies 98 GHz (referred to as 90 GHz as defined in ACT data products) and 150 GHz. The \texttt{daynight} maps are less noisy than the \texttt{night} maps, but exhibit larger beam systematics on small scales, due to daytime thermal expansion. We have checked that this choice does not bias our result.

\begin{figure}[!t]
    \centering
    \vspace{-0.5cm}
    \includegraphics[scale=0.8]{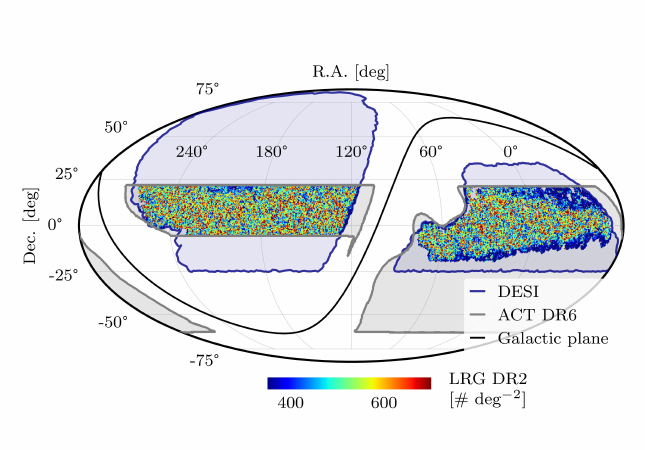}
    \vspace{-1cm}
    \caption{Angular density distribution of DESI DR2 LRGs before correction for observational completeness. The bluer region corresponds to area that are not fully complete. We show only the area common to DESI and ACT DR6. The nominal DESI survey (including its extension) will cover the entire blue shaded area, while the gray shaded region indicates the ACT DR6 mask used in the analysis as described in the text.}
    \label{fig:lrg_density}
\end{figure}

We apply the Planck \texttt{GAL070} mask to remove the region of the sky with high Galactic foreground emission, and we do not apply any tSZ cluster mask since we have verified that it does not change our results. The CMB pixel weight $W_{\rm CMB}(\boldsymbol{\theta})$ is set to be 1 where the noise is below 70 $\mu$K-arcmin for both the 90 and 150 GHz channels and 0 otherwise. The footprint of this total selection is shown in gray in \cref{fig:lrg_density}.

Finally, we equalize the filters $F_l$ such that 
\begin{equation} \label{eqn:norm_filters}
F_l^{150} = \dfrac{b_l^{150}}{b_l^{90}} F_l^{90}\,,
\end{equation}
whereby the two kSZ velocity bias parameters are expected to be essentially equal: $b_v^{90} \approx b_v^{150}$, and both velocity reconstructions can be compared together to perform a null test. The filters for the LRG case are displayed in \cref{fig:filters}. 

\subsection{DESI DR2}
The Dark Energy Spectroscopic Instrument (DESI) is a robotic, fiber-fed, highly multiplexed spectroscopic surveyor that operates on the Mayall 4-meter telescope at Kitt Peak National Observatory \citep{DESI2022,Silber2023}. DESI, which can obtain simultaneous spectra of almost 5000 objects over a $\sim3^o$ field \citep{Poppett2024,Miller2024}, is conducting an eight-year survey of about $17{,}000~\mathrm{deg}^2$ of the sky. The full survey will lead to 63 million spectroscopically-confirmed galaxies and quasars, compared to the initial forecasts of 39 million \citep{DESI2016II}. The scale and complexity of the DESI experiment necessitate a suite of supporting software pipelines and products to effectively exploit its data \citep{Guy2023,Myers2023,Schlafly2023}. 

We use the Luminous Red Galaxy (LRG) \cite{Zhou2023}, Emission Line Galaxy (ELG) \cite{Raichoor2023}, and Quasar (QSO) \cite{Chaussidon2023} samples from the DESI DR2 dataset, which comprises observations collected between 14 May 2021 and 9 April 2024 \cite{DESIDR2}. The corresponding cosmological interpretations from BAO measurements are presented in \cite{DESIDR2I, DESIDR2II}. Summary statistics for these samples, restricted to the area overlapping with ACT DR6, are reported in \cref{tab:desi_tracers}, and the observational footprint is shown in \cref{fig:lrg_density}. Together, these three tracers span a broad redshift range from $0.4 < z < 3.5$ and cover approximately  $5{,}500~\mathrm{deg}^2$ in common with ACT DR6.

\begin{table}[t!]
    \centering
    \caption{Summary statistics for the different tracers used in the analysis. All numbers refer to the common footprint between DESI and ACT DR6. The densities are crude estimates, while the effective redshifts $z_{\rm eff}^{\rm FKP}$\footnote{Afterwards, we identified an error in $z_{\rm eff}^{\rm FKP}$. The corrected values are \(0.794\), \(1.201\), and \(1.734\), respectively. This has no significant impact, since at linear scales, this small difference is absorbed by the linear bias.\vspace{-5mm}} are computed using FKP weighting. The completeness corresponds to the number of targets that are already observed.}
    \label{tab:desi_tracers}
    \vspace{1.5mm}
    \begin{tabular}{lccc}
    \toprule
     & LRG & ELG & QSO \\ \midrule \vspace{0.5mm}
    \# objects & 2,870,615 & 5,513,158 & 1,228,940 \\ \vspace{0.5mm}
     $z_{\rm min} - z_{\rm max}$ & 0.4 - 1.1 & 0.8 - 1.6 & 0.8 - 3.5 \\ \vspace{0.5mm}
     $z_{\rm eff}^{\rm FKP}$ & 0.734 & 1.170 & 1.667 \\ \vspace{0.5mm}
     $\bar{n}$ [Mpc$^{-3}$] & 1e-4 & 1.2e-4 & 0.68e-5 \\ \vspace{0.5mm}
     Area NGC [deg$^{2}$] & 2850 & 2864 & 3102 \\  \vspace{0.5mm}
     Area SGC [deg$^{2}$] & 2667 & 2679 & 3023 \\
     Completeness NGC [$\%$] & 89.6 & 59.1 & 97.6  \\
     Completeness SGC [$\%$] & 84.5 & 53.6 & 94.3  \\
    \bottomrule
    \end{tabular}
\end{table}

The clustering catalogs used in this analysis are described in detail in \cite{Ross2024, DESI2024II}, although they have minor differences. These choices are expected to be as close as possible to the fiducial ones for the upcoming DESI Key papers:
\begin{itemize}
\item We use the \texttt{\_zmb} version of the catalogs, which includes the conversion between the CMB rest frame and the observer rest frame.
\item We compute the FKP weights using different values of $P_0$ [$(\mathrm{Mpc}/h)^3$]: $5\times10^4$ for LRGs, $2\times10^4$ for ELGs, and $3\times10^4$ for QSOs.
\item We use the full ELG sample rather than the ELG\_LOP subsample that was used by default in the DR1 analysis. The ELG sample contains all the ELG observed with DESI, including ELG targets with the lowest priority in the observation that are preferentially objects with lower redshift ($z < 1.2$) \cite{Ross2024}. For DR2, this choice increases the number of ELGs by 19.5\%.
\item For imaging weights, we adopt a linear regression for QSOs following \cite{Chaussidon2024a}. For LRGs, we compute imaging weights in finer redshift bins ($\Delta z = 0.1$) and include a larger set of imaging maps in the regression, in particular the dust-related maps \texttt{EBV\_DIFF\_GR} and \texttt{EBV\_DIFF\_RZ}.
\end{itemize}

We apply the same weights to both the galaxy density and velocity fields\footnote{Data and random catalogs use identical weights, except for the additional completeness term (\texttt{FRAC\_TLOBS\_TILE}) applied to the randoms \cite{Ross2024}.}:
\begin{equation}
W^g = W^{\rm freq} = w_{\rm comp} \times w_{\rm sys} \times w_{\rm zfail} \times w_{\rm FKP}\,,
\end{equation}
which correct for observational completeness, fluctuations in target selection due to imaging systematics, redshift failure rates, and include the FKP weighting scheme \cite{Feldman1994}. In future work, we will use more optimal weighting scheme for the velocity field as derived in \cite{Howlett2019}. 

Note that, since we cross-correlate two observables that are not expected to share common angular systematics, the impact of imaging weights is expected to be minimal. As shown later, the large-scale modes of the dipole of the galaxy–velocity power spectrum do not exhibit excess of power for any of the three tracers, confirming this expectation (see \cref{fig:LRG_gv} and \cref{fig:ELG_QSO_gv}).

Finally, the power spectra are computed independently in the North (NGC) and South Galactic Cap (SGC). The consistency of the two measurements is assessed in \cref{app:ngc_vs_sgc}. They are then combined using a weighted average: $P(k) = w_1P_{\rm NGC}(k) + (1-w_1)P_{\rm SGC}(k)$ where $w_1$ is the ratio of the normalization factors of the two regions, following the DESI fiducial prescription \cite{DESI2024II}. This is done similarly for the different surrogate components to evaluate both the model and the covariance. The values of $w_1$ for each tracer are reported in \cref{tab:fiducial_parameters}. Owing to its higher observational completeness, the NGC contributes approximately $60\%$ of the total constraining power.

\section{Results} \label{sec:results}

\begin{figure*}[!htp]
    \centering
    \includegraphics[scale=0.9]{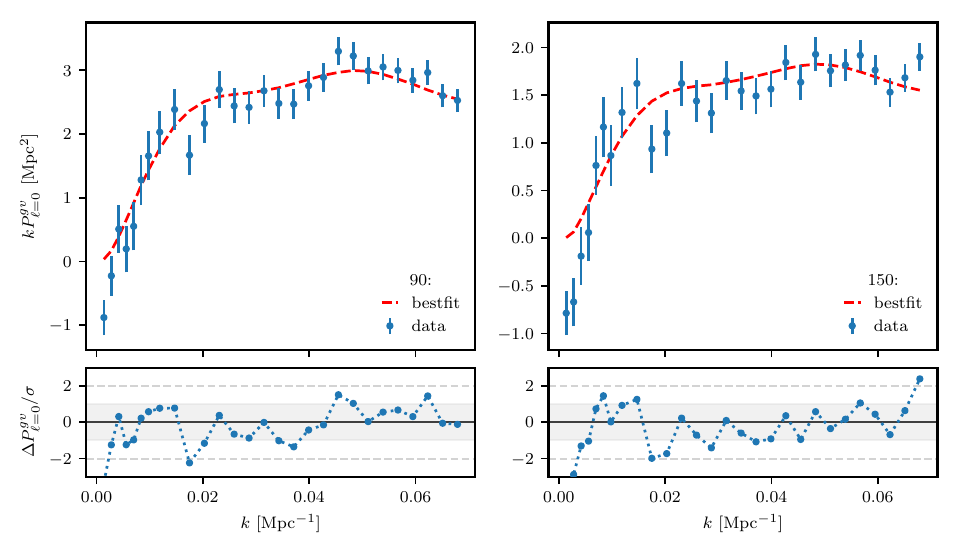}
    \caption{Monopole of the LRG galaxy-velocity power spectrum ($P_{\ell=0}^{gv}$), induced by the foregrounds, for the 90 (\textit{resp.} 150) GHz velocity reconstruction on the left (\textit{resp.} right). The best fits are displayed in red dashed lines and the best fit parameters are given in \cref{tab:lrg_pgv_ell0}. The residuals \ie (data - besfit) / errors are displayed in the bottom panel.}
    \label{fig:lrg_pgv_ell0}
\end{figure*}

The inferences were performed within \texttt{kszx}\footnote{\url{https://github.com/echaussidon/kszx/blob/main/kszx/Likelihood.py}.}. In particular, best-fit values are obtained with \texttt{scipy.optimize.minimize} \cite{Virtanen2020} using the \textit{Nelder-Mead} method, while the posteriors are sampled using a Monte Carlo Markov chain approach with \texttt{emcee} \cite{Foreman-Mackey2013} and displayed with \texttt{GetDist} \cite{Lewis2025}. Errors are given as the $1\sigma$ of the marginalized posteriors. 

\begin{table}[h!]
    \centering
    \caption{Fiducial parameters assumed to compute the covariance matrices. The first four parameters are fixed during the inference.}
    \label{tab:fiducial_parameters}
    \vspace{1.5mm}
    \begin{tabular}{lccc}
    \toprule
    params &  LRG & ELG & QSO \\
    \midrule
     $b_1$      & 2.1 & 1.3 & 2.48 \\
     $p$        & 1 & 1 & 1.4 \\
     $\Sigma_g$ & 0 & 0 & 0 \\
     $s_n$      & 0.99 & 1.05 & 0.99 \\
    \midrule
     $b_v^{90}$         & 0.224 & 0.10 & 0.24 \\
     $b_v^{150}$        & 0.228 & 0.14 & 0.21 \\ \vspace{0.3mm}
     $\Sigma_v$         & 7.5 &  7 & 11\\ \vspace{0.5mm}
     $\Sigma_{vv}^{v}$  & 7.5 &  7 & 11 \\ \vspace{0.5mm}
     $b_{\rm fg}^{90}$  & 1.82e-3  & 1.05e-4 & 2.12e-4 \\ \vspace{0.3mm}
     $b_{\rm fg}^{150}$ & 1.11e-3  & 2.41e-5 & 4.32e-5 \\
     $s_{nv}^{90}$      & 1.13 & 1.01 & 1 \\
     $s_{nv}^{150}$     & 1.09 & 1.01 & 1 \\
     $s_{nv}^{90x150}$  & 1.29 & 1.03 & 1 \\
     \midrule 
     $w_1$ & 0.57 & 0.59 & 0.59 \\
    \bottomrule
    \end{tabular}
\end{table}

Although the surrogate methodology allows the covariance matrix to be evaluated as a function of the fitted parameters, we fix the covariance during the inference to reduce the computational cost. The most time-consuming steps are the matrix inversion or its Cholesky decomposition, whose cost increases rapidly with the size of the data vector. The covariance matrices are therefore fixed at the fiducial parameter values listed in \cref{tab:fiducial_parameters}, which are obtained through an iterative fitting procedure on the data.

For guidance, \cref{tab:params_fixed} summarizes which parameters affect which power spectrum and whether they are fixed or fitted during inference. When held fixed, their values are taken from \cref{tab:fiducial_parameters}. Each parameter is described in \cref{sec:pipelines}, while \cref{sec:remarks} provides further details on the modeling choices, in particular on the different damping terms ($\Sigma$) or on the velocity noise rescaling terms ($s_{n,v}$).

\begin{table}[h!]
    \centering
    \caption{Summary of the parameters affecting each observable and whether they are fixed or fitted during the inferences. Note that $\Sigma_s$ is not mentioned as it fixed to 0 as described in \cref{sec:remarks}, $p$ is a tracer property, while $s_{n,0}$ only impact the covariance.}
    \label{tab:params_fixed}
    \vspace{1.5mm}
    \setlength{\tabcolsep}{12pt} 
    \begin{tabular}{lcc}
    \toprule
    Observable & fitted & fixed \\
    \midrule
    $P_{\ell=0}^{gv}$ & $b_{\rm fg}^{90}$, $b_{\rm fg}^{150}$& $\fnl$, $b_1$ \vspace{2mm} \\
    $P_{\ell=1}^{gv}$ & $\fnl$, $b_v^{90}$, $b_v^{150}$, $\Sigma_v$ & $b_1$, $b_{\rm fg}^{90}$, $b_{\rm fg}^{150}$ \vspace{2mm} \\
    $P_{\ell=1,\ell=1}^{vv}$ & \makecell{$b_v^{90}$, $b_v^{150}$, $\Sigma_{vv}^v$ \\ $s_{nv}^{90}$, $s_{nv}^{90x150}$, $s_{nv}^{150}$}& $b_{\rm fg}^{90}$, $b_{\rm fg}^{150}$ \\
    \bottomrule
    \end{tabular}
\end{table}

\subsection{Foregrounds contribution} \label{sec:lrg_foregrounds}
As discussed in \cref{sec:foreground_model}, foregrounds introduce an additional contribution approximately proportional to $b_{\rm fg}\,\delta_m$ in the reconstructed velocity field $\widehat{v}_r$. This induces a non-zero monopole (and other even multipoles) in the galaxy-velocity power spectrum, which then leaks into other multipoles --most notably the dipole-- through the window function convolution associated with the survey mask. We therefore first estimate the foreground contribution by inferring $b_{\rm fg}^{90}$ and $b_{\rm fg}^{150}$ from the monopole of the galaxy-velocity power spectrum.

The monopoles of the LRG galaxy–velocity power spectrum are shown in \cref{fig:lrg_pgv_ell0} for the 90~and~150\,GHz velocity reconstructions. They are detected with high significance, with a signal-to-noise ratio (SNR) of $\simeq 45$\footnote{The signal-to-noise ratio is defined as ${\rm SNR} = (m^{\rm T} C^{-1} m)^{1/2}$}, where $m$ denotes the best-fit theoretical model and $C$ the covariance matrix evaluated at the best-fit parameters. The covariance is corrected only by the Hartlap factor.. The best-fit values are given in \cref{tab:lrg_pgv_ell0}. The fit provides an acceptable description of the data, with a reduced chi-square value ($\chi^2_{\rm red}$) and a probability to exceed (PTE)\footnote{The reduced chi-square is defined as $\chi^2_{\rm red} = (d-m)^{\rm T} C^{-1} (d-m)/{\rm ndof}$, where $d$ is the data vector and ${\rm ndof}$ is the number of degrees of freedom, equal to the size of the data vector minus the number of fitted parameters. The covariance is corrected only by the Hartlap factor.} of $(1.22,0.126)$.

The two main potential sources of foreground contamination are the thermal Sunyaev-Zel'dovich (tSZ) effect and the cosmic infrared background (CIB). These two contributions have opposite signs: at 90 and 150 GHz, galaxies are anti-correlated with tSZ, whereas the CIB, which produces a temperature excess due to its emission, is positively correlated with the galaxies. 

\begin{table}[!t]
    \centering
    \caption{Best-fit results for the monopole of the LRG galaxy-velocity power spectra induced by the foregrounds: SNR~$=45.2$ and $\chi^2=66.0$ with 54 degrees of freedom leading to $\chi_{red}^2=1.22$ and PTE~$=0.13$. The true physical values of $b_{\rm fg}^{\rm true}$ are obtained by multiplying the measured values $b_{\rm fg}$ by $B$, as explained in \cref{eqn:true_bfg}.}   
    \vspace{1.5mm}
    \begin{tabular}{lccc}
    \toprule
     Params         & Best Fit &   Mean  & $\pm \sigma$     \\
     \midrule \vspace{0.2mm}
     $b_{\rm fg}^{90}$  & 1.82e-3  & 1.83e-3 & -4.37e-5/4.38e-5 \\ 
     $b_{\rm fg}^{150}$ & 1.11e-3  & 1.11e-3 & -3.60e-5/3.65e-5 \\
    \bottomrule
    \end{tabular}
    \label{tab:lrg_pgv_ell0}
\end{table}

The physical values of $b_{\rm fg}^{\rm true}$ are obtained by multiplying the measured $b_{\rm fg}$ by $B$ (see \cref{eqn:true_bfg}). The true values for the two frequencies are both negative, indicating that the dominant foreground contribution arises from the tSZ effect. Equally, the ratio $b_{\rm fg}^{90} / b_{\rm fg}^{150} \simeq 1.65$ closely matches the expected tSZ frequency scaling, $f_{\rm tSZ}(90)/f_{\rm tSZ}(150) \approx 1.68$\footnote{The frequency dependence of the tSZ effects is $f_{\rm tSZ}(\nu) = x\coth(x/2) - 4$ where $x = h\nu / k_B T_{\rm CMB}$ \cite{Carlstrom2002}.}, highlighting the dominance of the tSZ effect as the primary foreground contribution. 

On the very largest scales in \cref{fig:lrg_pgv_ell0}, we notice a small deviation from our model that has the opposite sign to the tSZ contribution. It can be a sign of a common residual systematics between galaxies and the ACT temperature maps (for example, galactic dust emission on large scales) or the expected small contributions from the Integrated Sachs-Wolfe (ISW) effect. To assess the impact of this residual, we include an additional $k^{-2}$ contribution in the foreground model. After accounting for its leakage into the dipole, this shifts $\fnl$ by less than $0.2\sigma$. A detailed modeling of the foreground monopole is left to future work, as it will be crucial for obtaining unbiased constraints on $\fnl$ given the substantial noise reduction expected from upcoming Simons Observatory data \cite{TheSimonsObservatoryCollaboration2025}.

A similar analysis is performed for the reconstructed velocity-galaxy correlation monopoles from ELG and QSO samples, but with significantly lower signal-to-noise ratios: SNR~$\sim 1.7$ for ELGs and SNR~$\sim 1.8$ for QSOs. The corresponding best-fit values are reported in \cref{tab:fiducial_parameters}. In these cases, the inferred foreground amplitudes are much smaller (after correcting for B) than those for LRGs, resulting in negligible leakage into the dipole compared to the statistical uncertainty of this observable.

\begin{figure}[!t]
    \centering 
    \includegraphics[width=1\linewidth]{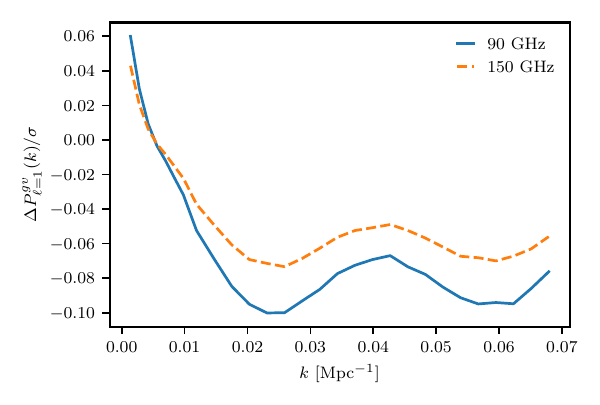}
    \caption{Residuals between the galaxy-velocity dipoles with and without the foreground contributions for the LRGs at the best fit values. This illustrates the small leakage of the monopole ($\ell=0$) induced by foregrounds into the dipole ($\ell=1$) because of the window function.}
    \label{fig:monopole_leakage}
\end{figure}

A non-zero monopole leaks into the dipole through the survey window function. To account for this effect and avoid biasing the remaining fits, we fix the foreground bias parameters $b_{\rm fg}$ to their best-fit values inferred from the monopole. As shown in \cref{fig:monopole_leakage}, this leakage affects primarily the small-scale dipole modes and remains small below $\sim 10\%$ of the statistical uncertainties for the LRGs.

Overall, this demonstrates that while foregrounds generate a highly significant monopole signal, their induced leakage into the dipole remains subdominant to the statistical errors and can be robustly controlled within our modeling framework. Once included, the rest of the analysis is unbiased. A similar conclusion applies to the velocity-velocity power spectrum, where the foreground contribution is treated consistently.

\subsection{Luminous Red Galaxies} \label{sec:lrg_analysis}
\begin{figure*}[!hbtp]
    \centering
    \subfloat[Dipole of the LRG galaxy-velocity power spectrum ($P_{\ell=1}^{gv}$) detected at $\sim 17\sigma$. \label{fig:LRG_gv}]{\includegraphics[scale=0.9]{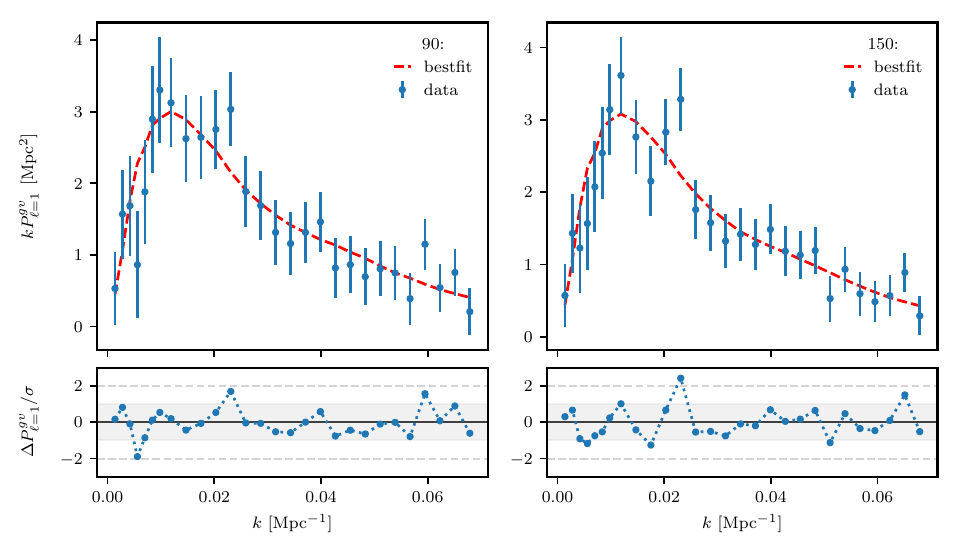}} \\ 
    \vspace{0.7cm}
    \includegraphics[width=\textwidth]{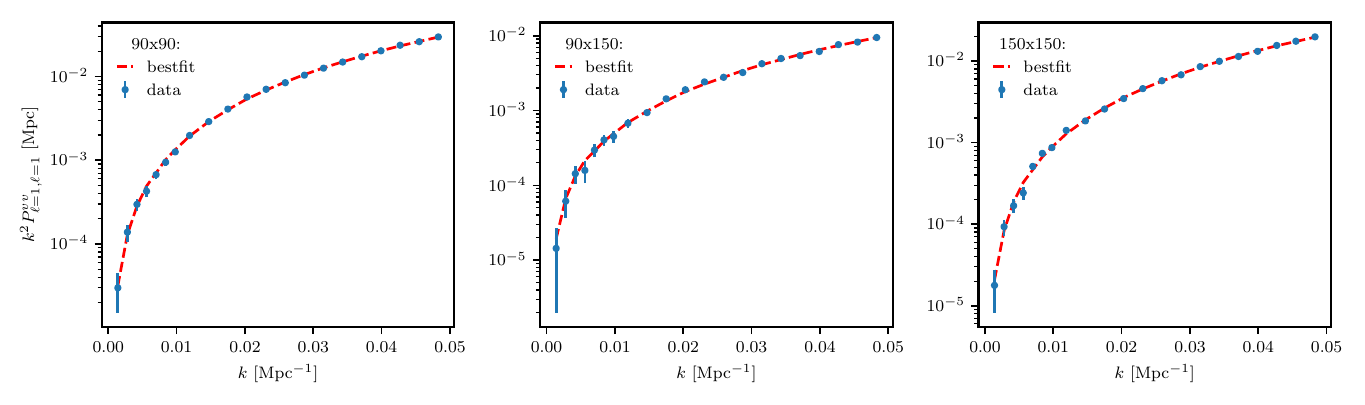} \\
    \vspace{-1.33cm}
    \subfloat[LRG velocity-velocity power spectrum ($P_{\ell=1,\ell=1}^{vv}$) with velocity reconstruction noise and foregrounds (top) with the best fit velocity reconstruction noise and foregrounds subtracted (bottom), detected at $\sim 3.1\sigma$.\label{fig:LRG_vv}]{\includegraphics[width=\textwidth]{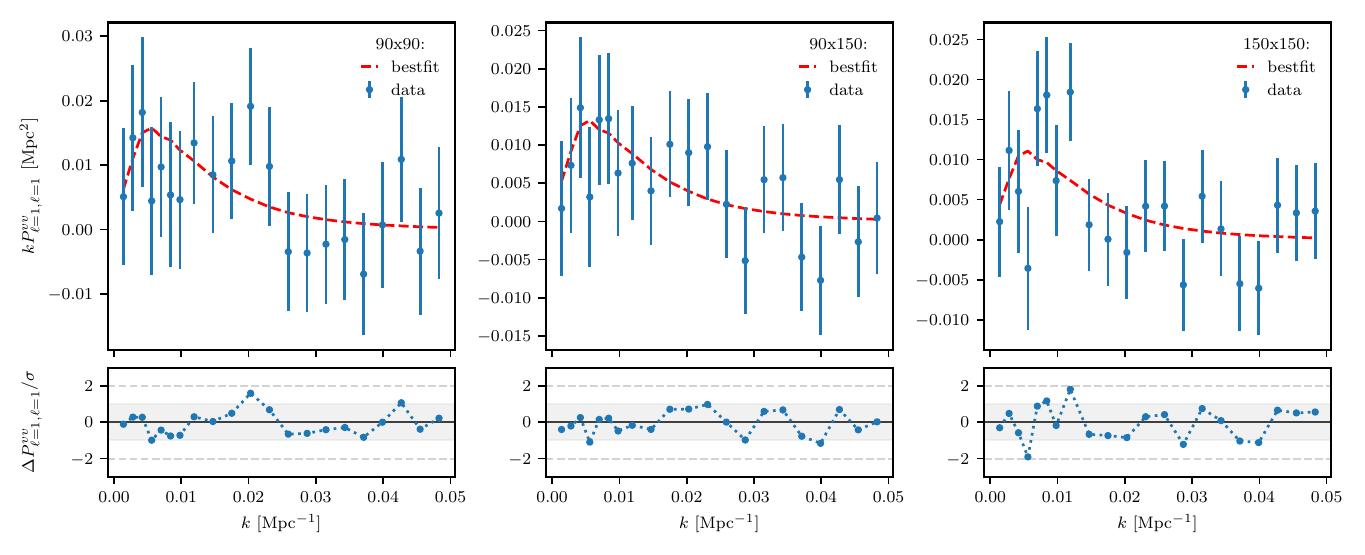}}
    \caption{Blue points are the data and red dashed lines are the best fit that is given in \cref{tab:lrg_gv_vv} while the posterior is displayed in \cref{fig:LRG_mcmc_gv}. (a) Left (Right) is for the velocity reconstructed with the 90~(150)~GHz. (b) Left, Middle, Right is for the cross-correlation between the 90$\times$90, 90$\times$150, 150$\times$150 reconstructed velocities.}
    \vspace{0.3cm}
\end{figure*}

\cref{fig:LRG_gv} and \cref{fig:LRG_vv} show the main results of this analysis (blue dotted points): the dipole of the LRG galaxy-velocity power spectrum detected at $\sim 17\sigma$, the highest significance to date, and the LRG velocity-velocity power spectrum detected for the first time at $\sim 3.1 \sigma$. In these figures, the red dashed lines show the best fits from the joint analysis detailed below. The best-fit values are reported in the first rows of \cref{tab:lrg_gv_vv}. Both spectra are shown down to $k_{\rm min} = 0.0014~\rm{Mpc}^{-1}$ without exhibiting any excess of power at large scales commonly observed in the galaxy-galaxy power spectrum measurements \cite{Chaussidon2024a}.

\begin{table}[!h]
    \centering
    \caption{Best-fit results for the dipole of the LRG galaxy-velocity power spectra (top rows), for the velocity-velocity power spectra (middle rows) and for the combination of the two (bottom rows), with in the same order: SNR~$=17.05$ and $\chi^2=37.76$ with 52 degrees of freedom leading to $\chi_{red}^2=0.726$ and PTE~$=0.93$ (top rows), SNR~$=3.11$ and $\chi^2=51.118$ with 57 degrees of freedom leading to $\chi_{red}^2=0.897$ and PTE~$=0.69$ (middle rows), and SNR~$=17.9$ and $\chi^2=88.374$ with 111 degrees of freedom leading to $\chi_{red}^2=0.796$ and PTE~$=0.94$ (bottom rows). Posteriors are displayed in \cref{fig:LRG_mcmc_vv} and in \cref{fig:LRG_mcmc_gv}.}
    \vspace{1.5mm}
    \begin{tabular}{lccc}
    \toprule
     Params           & Best Fit & Mean  & $\pm \sigma$     \\
     \midrule \multicolumn{4}{c}{$P_{\ell=1}^{gv}$} \\ \midrule
     $\fnl$      &    26.5  & 33.9   & -56.8/56.7       \\
     $b_v^{90}$  &     0.224 &  0.219 & -0.0229/0.0228 \\
     $b_v^{150}$ &     0.228 &  0.224 & -0.0216/0.0216 \\
     $\Sigma_v$  &     8.17  &  7.37  & -2.36/2.29    \\
     \midrule \multicolumn{4}{c}{$P_{\ell=1,\ell=1}^{vv}$} \\ \midrule
     $b_v^{90}$       & 0.298   & 0.273 & -0.0786/0.0778   \\
     $b_v^{150}$      & 0.248   & 0.232 & -0.0639/0.0643   \\
     $\Sigma_v^{vv}$  & 12.3    & 21.0  & -14.6/15.0       \\
     $s_{nv}^{90}$    & 1.13    & 1.13  & -0.00705/0.00703 \\
     $s_{nv}^{150}$   & 1.09    & 1.09   & -0.00559/0.00561 \\
     $s_{nv}^{90x150}$& 1.28    & 1.29  & -0.0172/0.0172   \\ 
     \midrule \multicolumn{4}{c}{$P_{\ell=1}^{gv} + P_{\ell=1,\ell=1}^{vv}$} \\ \midrule 
     $\fnl$           & 8.5    & 11.7 & -43.5/43.4       \\
     $b_v^{90}$       & 0.233    & 0.229 & -0.0223/0.0224   \\
     $b_v^{150}$      & 0.237    & 0.233 & -0.0199/0.0199   \\
     $\Sigma_v$       & 8.56     & 7.91  & -1.95/2.00      \\
     $\Sigma_v^{vv}$  & 1.02e-05 & 17.8  & -12.9/13.3       \\
     $s_{nv}^{90}$    & 1.13     & 1.13  & -0.00690/0.00691 \\
     $s_{nv}^{150}$   & 1.09     & 1.09  & -0.00544/0.00545 \\
     $s_{nv}^{90x150}$& 1.28     & 1.29  & -0.0170/0.0171 \\
    \bottomrule
    \end{tabular}
    \label{tab:lrg_gv_vv}
\end{table}

First, the two reconstructed velocity fields from the 90 and 150 GHz channels are highly correlated, as they trace the same underlying velocity field. However, their reconstruction noises differ, so jointly fitting the observables derived from both reconstructions improves the overall constraints.

The velocity–velocity power spectra are used to constrain the three velocity reconstruction noise parameters $s_{nv}$. The best-fit values are reported in the middle rows of \cref{tab:lrg_gv_vv}. We find that the bootstrap strategy used to generate the surrogate noise slightly underestimates the reconstruction noise, as indicated by $s_{nv}^{90}, s_{nv}^{150} > 1$. This likely reflects limitations of the bootstrap sampling, since for LRGs the inverse number density is small compared to the corresponding $C_\ell$ of the same tracer. For tracers such as ELGs and QSOs, which are in a more shot-noise-dominated regime, we do not observe this effect. One could imagine using a \emph{block bootstrap} to circumvent this issue. We additionally introduce a third reconstruction noise parameter, $s_{nv}^{90\times150}$, as the cross-correlation exhibits a higher noise level than predicted by the surrogates. These excess noise contributions are consistently propagated into the covariance matrix, as described in \cref{sec:remarks}.

\begin{figure}[!tp]
    \centering
    \includegraphics[width=1\linewidth]{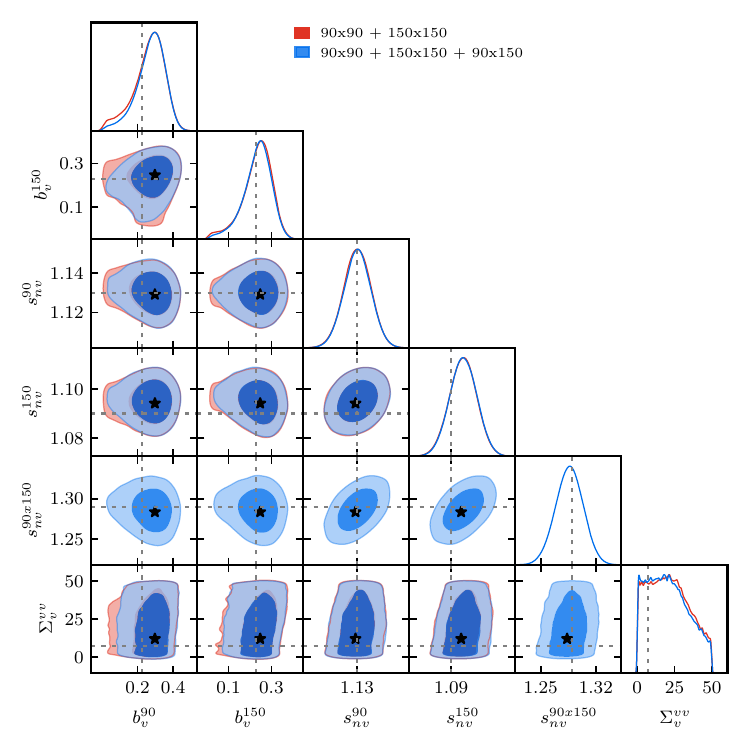}
    \caption{Posteriors obtained with LRG $P_{\ell=1,\ell=1}^{vv}$ with (\textit{resp.} without) the inclusion of the cross-correlation between the 90 GHz velocity and the 150 GHz velocity in blue (\textit{resp.} red). Best fits (black stars) for the complete case are given in \cref{tab:lrg_gv_vv}. The gray dashed lines are the fiducial parameter values adopted for the covariance.}
    \label{fig:LRG_mcmc_vv}
\end{figure}

\begin{figure}[!p]
    \centering
    \includegraphics[width=1\linewidth]{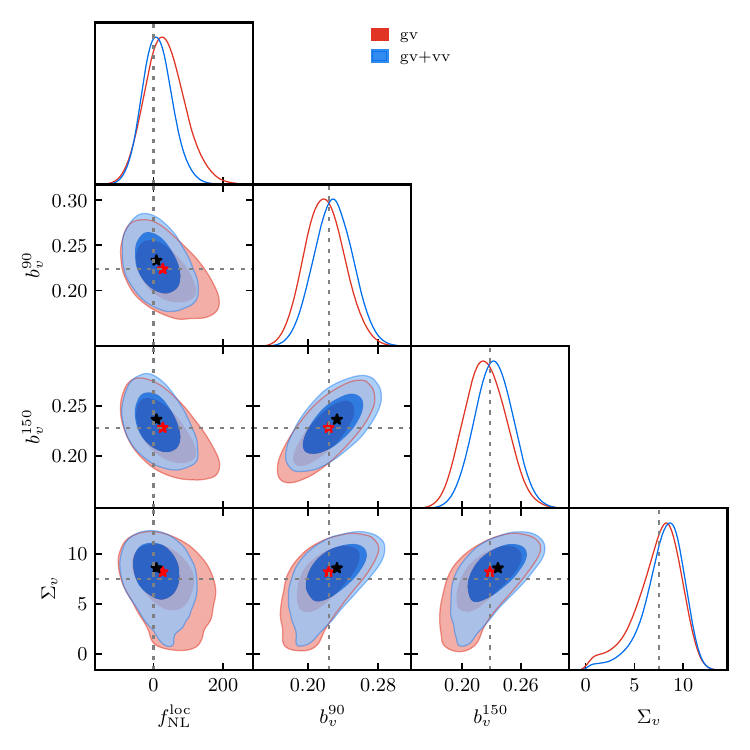}
    \caption{Comparison of the posteriors obtained with  $P_{\ell=1}^{gv}$ (red) with red star the best fits and the ones obtained with $P_{\ell=1}^{gv} + P_{\ell=1,\ell=1}^{vv}$ (blue) with black star the best fits. Best fits are given in \cref{tab:lrg_gv_vv}. The gray dashed lines are the fiducial parameter values adopted for the covariance. The inclusion of $P_{\ell=1,\ell=1}^{vv}$ reduce by $23\%$ the uncertainty on $\fnl$ and produce more gaussian posteriors.}
    \label{fig:LRG_mcmc_gv}
\end{figure}

The reconstruction noise terms are constant as a function of $k$ and dominate the measured velocity–velocity power spectra, with an overall signal-to-noise ratio of SNR~$\sim 123$, as shown in the top panels of \cref{fig:LRG_vv}. To assess the detection significance of the velocity signal itself, we subtract the best-fit noise and foreground contributions from the measured $P_{\ell=1,\ell=1}^{vv}$, yielding the residuals shown in the lower panels of \cref{fig:LRG_vv}. Including the $90\times150$ cross-correlation, detected at $2.54\sigma$, increases the overall detection significance of the velocity–velocity power spectrum from $2.99\sigma$ to $3.11\sigma$, and slightly improves the resulting parameter constraints, as shown in \cref{fig:LRG_mcmc_vv}. We tested that the SNR remains stable under changes to the fiducial value used to compute the covariance matrix. For comparison with the monopole measurement $P_{\ell=0,\ell=0}^{vv}$, see \cref{app:ell00vsell11}.

Finally, we perform a joint fit using both $P_{\ell=1}^{gv}$ and $P_{\ell=1,\ell=1}^{vv}$ to constrain $\fnl$. The corresponding correlation matrix is shown in the appendix in \cref{fig:corr_example}. The posterior distributions are displayed in blue in \cref{fig:LRG_mcmc_gv}, with the best-fit values reported in the last rows of \cref{tab:lrg_gv_vv}. For comparison, we also show the joint fit of the two $P_{\ell=1}^{gv}$ measurements from the 90 GHz and 150 GHz velocity reconstructions, shown by the red contours, and the corresponding best-fit values in the top row of the table.

Because $P_{\ell=1}^{gv}$ is detected with a higher signal-to-noise ratio (SNR$~\sim 17$) than $P_{\ell=1,\ell=1}^{vv}$, the resulting constraints on $b_v$ are tighter. The two sets of constraints are fully consistent, although $P_{\ell=1,\ell=1}^{vv}$ shows a mild preference for slightly larger values of $b_v$. The value of $b_v$ is discussed in \cref{sec:low_bv}.

All fits provide a good description of the data, with $(\chi^2_{\rm red},{\rm PTE})$ values of $(0.726\seliminline{,}0.93)$, $(0.897\seliminline{,}0.69)$, and $(0.796\seliminline{,}0.94)$ for $P_{\ell=1}^{gv}$, $P_{\ell=1,\ell=1}^{vv}$, and the joint fit $P_{\ell=1}^{gv} + P_{\ell=1,\ell=1}^{vv}$, respectively. As consistency checks, we compare the NGC and SGC fits, as shown in \cref{app:ngc_vs_sgc}, and perform a null test by computing $P_{\ell=1}^{gv}$ using the difference of the two velocity fields $(\widehat{v}_r^{90} - \widehat{v}_r^{150})$, as presented in \cref{app:null_test}.

In the joint fit, we use two distinct damping parameters, $\Sigma_v$ and $\Sigma_v^{vv}$, as explained in \cref{sec:remarks}. This is because we fix $\Sigma_g = 0$ in $P_{\ell=1}^{gv}$, such that $\Sigma_v$ effectively accounts for the damping contributions from both the velocity and galaxy fields.

Including $P_{\ell=1,\ell=1}^{vv}$ leads to an overall improvement in the constraints compared to using $P_{\ell=1}^{gv}$ alone. In particular, we find a $\sim 23\%$ reduction in the uncertainty on $\fnl$, despite the fact that the velocity–velocity power spectrum itself carries no direct information on $\fnl$. This improvement arises from tighter constraints on $b_v$ that break the natural degeneracy between $\fnl$, $b_v$, and $\Sigma_v$, as illustrated in \cref{fig:LRG_mcmc_gv}.

\begin{figure}[!tp]
    \centering
    \includegraphics[width=1\linewidth]{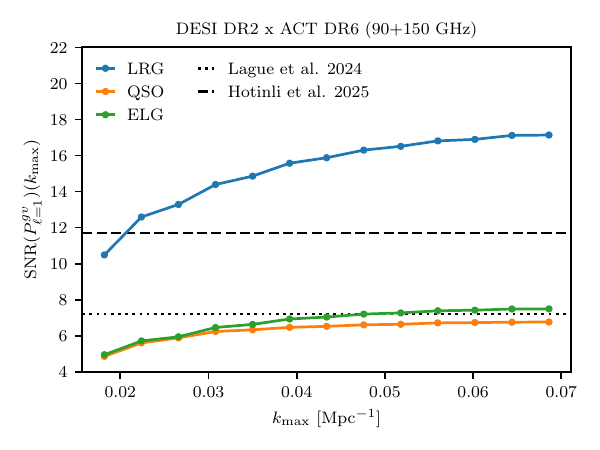}
    \caption{Evolution of the signal-to-noise ratio as a function of $k_{\rm max}$ with $k_{\rm min}=0.0014$ for the different tracers. These SNR are only for the dipole of the galaxy-velocity power spectra. Dotted and dashed black lines were previous study using either CMASS galaxies from BOSS \cite{Lague2024} or photometric LRG sample from DESI targeting \cite{Hotinli2025} in combination with ACT DR5.}
    \label{fig:SNR}
\end{figure}

The cumulative SNR as a function of $k_{\rm max}$ is shown in \cref{fig:SNR}. It steadily increases as intermediate scales are included, before reaching a plateau where the velocity reconstruction noise becomes dominant. The onset of this plateau defines the maximum wavenumber, $k_{\rm max}$, used in this analysis, since no additional velocity information can be reliably extracted beyond this scale. For comparison, the dotted and dashed black lines indicate the SNR obtained in previous analyses using either the CMASS galaxy sample from BOSS using $k_{\rm max} = 0.03~\rm{Mpc}^{-1}$ \cite{Lague2024} or photometric LRGs from the DESI targeting catalogs using $k_{\rm max} = 0.018~\rm{Mpc}^{-1}$ \cite{Hotinli2025} in combination with ACT DR5. The substantial improvement observed here demonstrates the enhanced constraining power provided by the DESI DR2 LRG spectroscopic samples.

Finally, we attempt to detect the signal in the octopole ($\ell = 3$), but find no significant detection (SNR~$= 0.74$), as shown in \cref{app:octopole}. However, the octopole does not exhibit any large systematic effects.

\subsection{Emission Line Galaxies \& Quasars} 
This section presents the two other important results of this analysis. The first detection of the kSZ effect using the Emission Line Galaxies (ELG) at $8.3 \sigma$ and the Quasars (QSO) at $6.8\sigma$. 

Similarly, we start by fitting the foregrounds (see \cref{sec:lrg_foregrounds}) and $P_{\ell=1,\ell=1}^{vv}$ to obtain the velocity reconstruction noises $s_{nv}$ and so the correct covariance matrix. We do not report any detection of the velocity-velocity signal for both tracers.

\begin{figure*}[!hbtp]
    \vspace{1.5cm}
    \centering
    \subfloat[Dipole of the ELG galaxy-velocity power spectrum ($P_{\ell=1}^{gv}$) detected at $\sim 8.3\sigma$.]{\label{fig:ELG_gv}\includegraphics[scale=0.9]{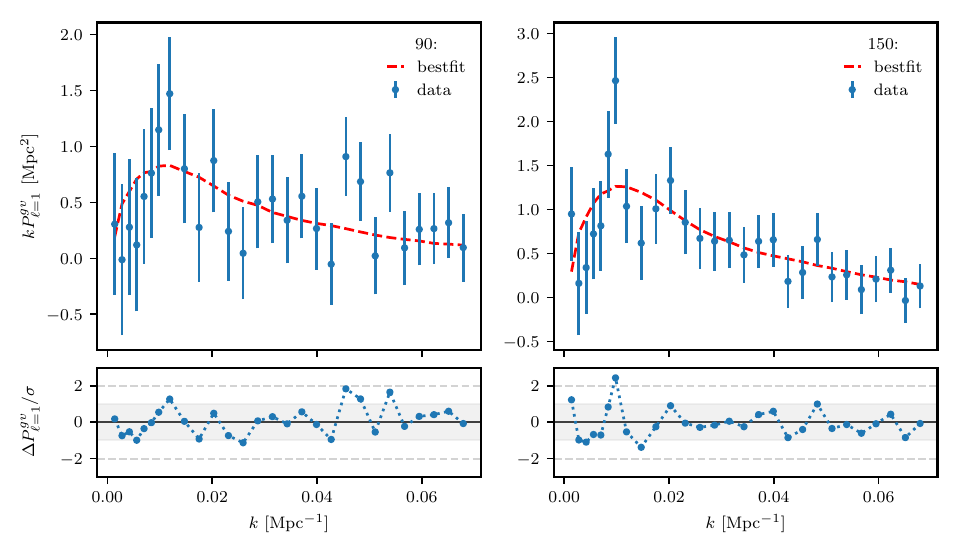}} \\
    \subfloat[Dipole of the QSO galaxy-velocity power spectrum ($P_{\ell=1}^{gv}$) detected at $\sim 6.8\sigma$.]{\label{fig:QSO_gv}\includegraphics[scale=0.9]{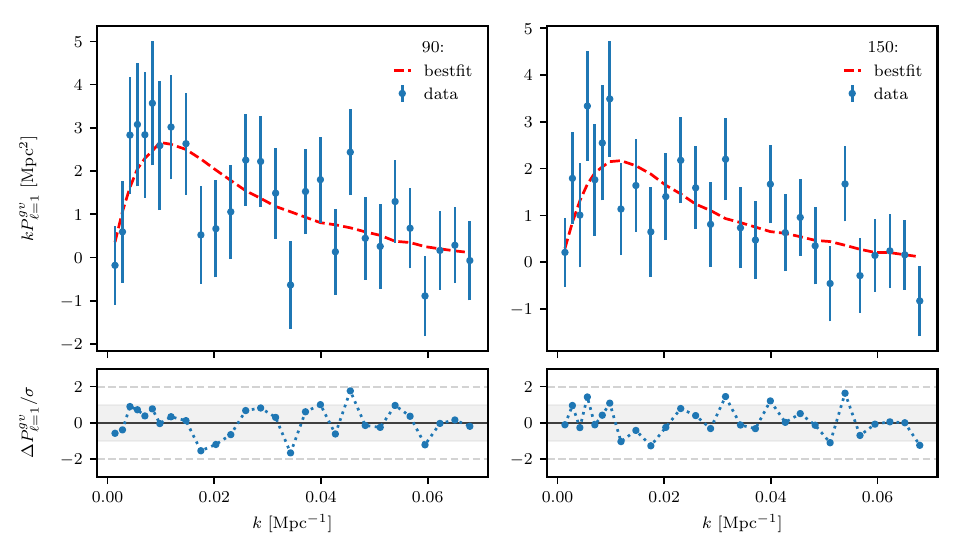}}
    \caption{Similar than \cref{fig:LRG_gv} but for ELG (a) and QSO (b). The best fits are given in \cref{tab:elg_qso_gv} and the posteriors are displayed in \cref{fig:ELG_QSO_mcmc_gv}.}
    \vspace{1cm}
    \label{fig:ELG_QSO_gv}
    \newpage
\end{figure*}

\begin{figure*}[!htp]
    \centering
    \subfloat[ELG]{\includegraphics[scale=0.65]{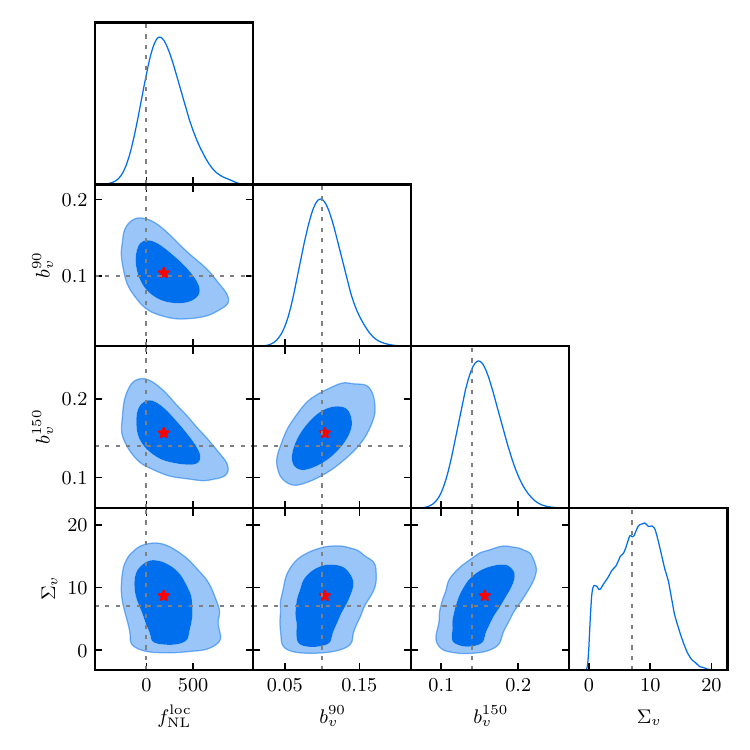}}\hfill
    \subfloat[QSO]{\includegraphics[scale=0.65]{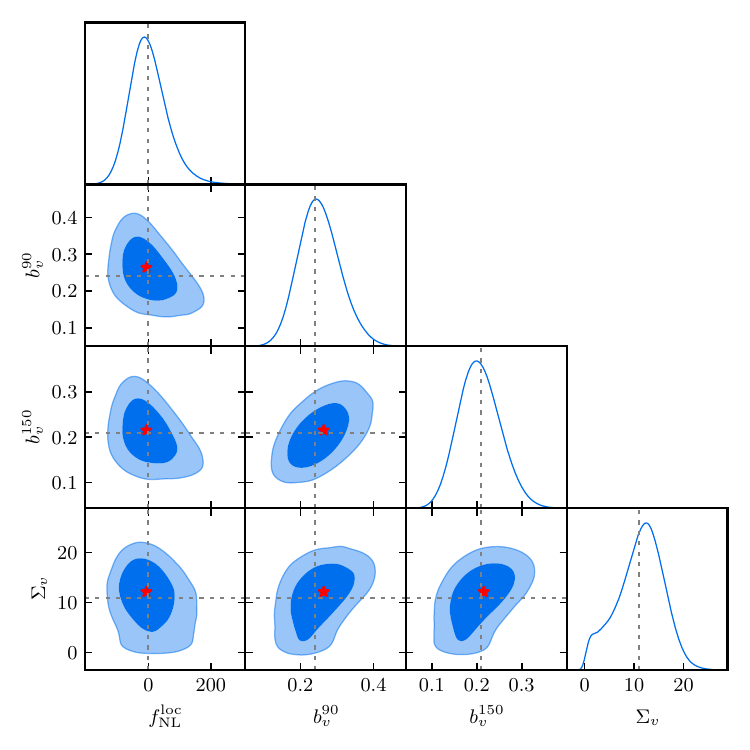}}
    \caption{Posteriors obtained from the dipole of the ELG (left) and QSO (right) galaxy-velocity power spectrum. The two posteriors are independent. The best fits (red stars) are given in \cref{tab:elg_qso_gv}. The gray dashed lines are the fiducial parameter values adopted for the covariance.}
    \label{fig:ELG_QSO_mcmc_gv}
\end{figure*}

The ELG and QSO galaxy-velocity power spectra are shown in \cref{fig:ELG_QSO_gv}, while the posteriors are shown in \cref{fig:ELG_QSO_mcmc_gv} and the best fits in \cref{tab:elg_qso_gv}. Again, the fits provide a good description of the data with ($\chi^2_{\rm red}$, PTE) equal to $(1.061, 0.36)$ for ELGs  and $(0.707, 0.95)$ for QSOs. The cumulative SNR as a function of $k_{\rm max}$ for the ELGs and QSOs are displayed in \cref{fig:SNR}. The interpretation of $b_v$ is discussed in \cref{sec:low_bv}, in particular, the origin of the difference between the ELG measurement and those of LRGs and QSOs.

\begin{table}[b!]
    \centering
    \caption{Best-fit results for the dipole of the ELG galaxy-velocity power spectra (top rows) and for the QSOs (bottom rows) with SNR~$=8.3$ and $\chi^2=55.15$ with 52 degrees of freedom leading to $\chi_{red}^2=1.061$ and PTE~$=0.36$ (top rows), and SNR~$=6.8$ and $\chi^2=36.7$ with 52 degrees of freedom leading to $\chi_{red}^2=0.707$ and PTE~$=0.95$ (bottom rows). Posteriors are displayed in \cref{fig:ELG_QSO_mcmc_gv}.}
    \vspace{1.5mm}
    \begin{tabular}{lccc}
    \toprule
     Params      &   Best Fit &    Mean & $\pm \sigma$  \\ \midrule
     \multicolumn{4}{c}{ELG} \\ 
     \midrule
     $\fnl$      &   186     & 207     & -221/225       \\
     $b_v^{90}$  &     0.104 &   0.102 & -0.0269/0.0270 \\
     $b_v^{150}$ &     0.156 &   0.153 & -0.0259/0.0261 \\
     $\Sigma^v$  &     8.67  &   7.82  & -4.57/4.18     \\
    \midrule 
    \multicolumn{4}{c}{QSO}\\
    \midrule
     $\fnl$      &   -6.69   & 1.09    & -58.4/57.7     \\
     $b_v^{90}$  &     0.264 &   0.252 & -0.0562/0.0566 \\
     $b_v^{150}$ &     0.215 &   0.206 & -0.0467/0.0469 \\
     $\Sigma_v$  &    12.1   &  11.2   & -5.02/4.62     \\
    \bottomrule
    \end{tabular}
    \label{tab:elg_qso_gv}
\end{table}

Despite the higher signal-to-noise ratio observed for the ELG sample, the constraints on $\fnl$ are weaker. This is because the linear bias $b_1$ of ELGs is relatively close to unity, which results in a value of the PNG-induced bias parameter $b_\phi$ close to zero, thereby reducing the sensitivity of this tracer to primordial non-Gaussianity.

For the quasar sample, despite a lower detection significance compared to LRGs, the uncertainties on the large-scale modes of the galaxy–velocity power spectrum remain competitive for constraining $\fnl$, due to the very large cosmological volume probed ($0.8 < z < 3.5$). Moreover, the PNG response parameter $b_\phi$ is comparable for LRGs and QSOs.

\subsection{Different tracers combined} \label{sec:full_analysis}
\begin{figure}[!bp]
    \centering
    \includegraphics[width=0.8\linewidth]{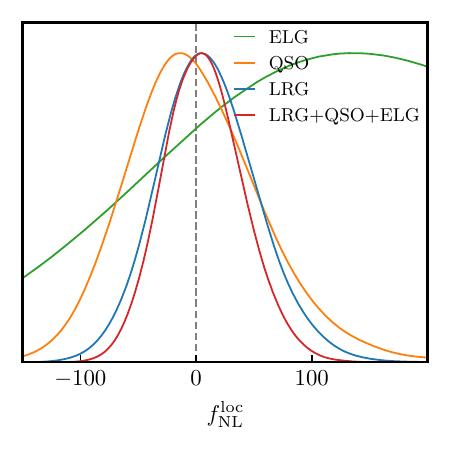}
    \caption{$\fnl$ posteriors from the different tracers (ELG in green, QSO in orange and LRG in blue) and for the combined analysis in red. LRG and QSO two most constraining tracers are in very good agreement. Best fits are given in \cref{tab:lrg_gv_vv}, \cref{tab:elg_qso_gv} and in \cref{tab:lrg+elg+qso} for the combination.}
    \label{fig:combined_tracers_fnl}
\end{figure}

The combined posterior for $\fnl$ is shown in \cref{fig:combined_tracers_fnl}, with the corresponding best-fit values reported in \cref{tab:lrg+elg+qso}. The full corner plot is provided in the appendix in \cref{fig:combined_tracers_full}. The inclusion of ELGs has a negligible impact relative to the combination of LRGs and QSOs; nevertheless, it is retained here since it does not bias the results nor complicate the analysis. This joint analysis yields the tightest constraints on $\fnl$ to date obtained using the reconstructed velocity fields:
\begin{equation}
\fnl=\begin{cases} \vspace{1mm}
8.5 _{-43.5}^{+43.4} & \text { LRG } \\ \vspace{1mm}
186_{-221}^{+225} & \text { ELG } \\ \vspace{1mm}
-6.69_{-58.4}^{+57.7} & \text { QSO } \\ \vspace{1mm} 
15.9_{-34.4}^{+34.6} & \text { LRG + ELG + QSO } \\
\end{cases},
\end{equation}
at 68\% confidence.

\begin{table}[!th]
    \centering
    \caption{Best-fit results for the combination of the different observables presented in this analysis with SNR~$=20.8$ and $\chi^2=177.5$ with 217 degrees of freedom leading to $\chi_{red}^2=0.818$ and PTE~$=0.97$. $\fnl$ is displayed in \cref{fig:combined_tracers_fnl}.}
    \label{tab:lrg+elg+qso}
    \vspace{1.5mm}
    \begin{tabular}{llccc}
    \toprule
             & Params            &Best Fit&  Mean &   $\pm \sigma$   \\ \midrule
             & $\fnl$            & 15.9   & 6.68  & -34.4/34.6       \\ \midrule
             & $b_v^{90}$        & 0.221  & 0.226 & -0.0217/0.0216   \\
             & $b_v^{150}$       & 0.225  & 0.230 & -0.0193/0.0192   \\
\hspace{9mm} & $\Sigma_v$        & 7.56   & 8.00  & -1.95/1.97       \\
     LRG     & $\Sigma_v^{vv}$   & 20.3   & 16.3  & -11.6/12.1       \\
             & $s_{nv}^{90}$     & 1.13   & 1.13  & -0.00692/0.00701 \\
             & $s_{nv}^{150}$    & 1.1    & 1.09  & -0.00551/0.00553 \\
             & $s_{nv}^{90x150}$ & 1.29   & 1.29  & -0.0169/0.0169   \\ \midrule
             & $b_v^{90}$        & 0.227  & 0.254 & -0.0529/0.0528   \\
     QSO     & $b_v^{150}$       & 0.176  & 0.207 & -0.0441/0.0444   \\
             & $\Sigma_v$        & 7.57   & 11.3  & -4.94/4.58       \\ \midrule
             & $b_v^{90}$        & 0.115  & 0.114 & -0.0254/0.0255   \\
     ELG     & $b_v^{150}$       & 0.167  & 0.167 & -0.0245/0.0244   \\
             & $\Sigma_v$        & 9.34   & 8.75  & -4.31/3.90       \\
    \bottomrule
    \end{tabular}  
\end{table}

In this analysis, since the linear bias $b_1$ is degenerate with the amplitude of the kSZ signal $b_v$, we fix its value to that measured from the monopole of the galaxy power spectrum. Consequently, we do not propagate uncertainties in $b_1$ into the constraint on $\fnl$\footnote{Running the inference with a gaussian prior on $b_1$ increases the errors on $\fnl$ by a percent.}. In the future, this will be naturally accounted for, as we plan to perform a combined analysis of the galaxy–galaxy, galaxy–velocity, and velocity–velocity power spectra once the blinding policy of DESI allows it.

\subsection{Lower kSZ signal than expected} \label{sec:low_bv}
As discussed in \cref{sec:velocity_bias}, the reconstructed radial velocity field $\widehat{v}_r$ provides a biased estimate of the true velocity field. In particular, we marginalize over the kSZ velocity bias parameter $b_v$, since the overall amplitude of the kSZ signal is not known \emph{a priori} \cite{Giri2022}. The parameter $b_v$ is defined as the ratio between the integrated true and fiducial galaxy–electron power spectra:
\begin{equation*}
b_v(\chi) =
\frac{\int \mathrm{d}^2\mathbf{L}\,b_L\,F_L\, P_{ge}^{\rm true}(L/\chi)}
{\int \mathrm{d}^2\mathbf{L}\,b_L\,F_L\,P_{ge}^{\rm fid}(L/\chi)}\,.
\end{equation*}
This integration is dominated by small scales, as large-scale modes are suppressed by the filter $F_L$. In the ideal case where the small-scale galaxy and electron physics are perfectly modeled, one would expect $b_v = 1$.

To compute the fiducial galaxy–electron power spectrum $P_{ge}^{\rm fid}$, we follow the halo-model prescription implemented in \texttt{hmvec}, which adopts a halo occupation distribution (HOD) based on abundance matching to the mean galaxy population, as described in Appendix B of \cite{Smith2018}. This modeling performs well for the LRG and QSO samples \cite{Yuan2023}. However, abundance matching provides a poor description of the ELG sample, as discussed in \cite{Rocher2023}. As a result, when computing $P_{ge}^{\rm fid}$ for ELGs, the HOD we adopt yields to a linear bias of $b_1 \sim 2.3$ rather than the nominal value $b_1 \simeq 1.3$, which leads to a bias towards a low value of $b_v$. This does not bias our results, as we marginalize over $b_v$. A more accurate description of the ELG HOD will be implemented in future work.

Although the HOD model adopted here does not perfectly match the DESI best-fit galaxy properties, it provides a reasonably accurate description of the galaxy contribution to $P_{ge}$, since the dominant effect at leading order is the overall signal amplitude set by $b_1$. We therefore consistently find low values of the kSZ signal amplitude across a wide redshift range. This result supports —and extends to higher redshifts, enabled by the inclusion of ELGs and QSOs— previous findings \cite{Hadzhiyska2025, Guachalla2025,Hadzhiyska2025a} 
reflecting a modeling error in the assumed electron profile, potentially explained by enhanced baryonic feedback compared to Battaglia \cite{Battaglia2016}. We note that our halo-model prescription for the electron density profile adopts the Battaglia AGN profile, implemented in the way described in the appendix B of \cite{Smith2018}, where the baryon fraction reaches the cosmological baryon fraction at the halo virial radius. Further tests could be performed by comparing to a wider range of feedback models from hydrodynamical simulations, such as \textsc{Illustris} \cite{Nelson2018} or \textsc{Flamingo} \cite{Schaye2023}, and this will be the subject of future work. 

In \cite{Hotinli2026}, we compare this analysis in more detail with the one performed using DESI LRG targets in \cite{Hotinli2025}, and we show agreement between the two.

In future work, we plan to refine both the galaxy and electron components of the model in order to adopt a more accurate description of $P_{ge}$ and construct more optimal filters $F_L$, to improve the overall performance of our estimators. This, together with a comparison with recent stacking measurements \cite{Hadzhiyska2025, Guachalla2025}, will be investigated in \cite{Chen2026}.

\subsection{Large scale noise: galaxies vs. kSZ velocities}
\begin{figure}[!tp]
    \centering
    \includegraphics[width=1\linewidth]{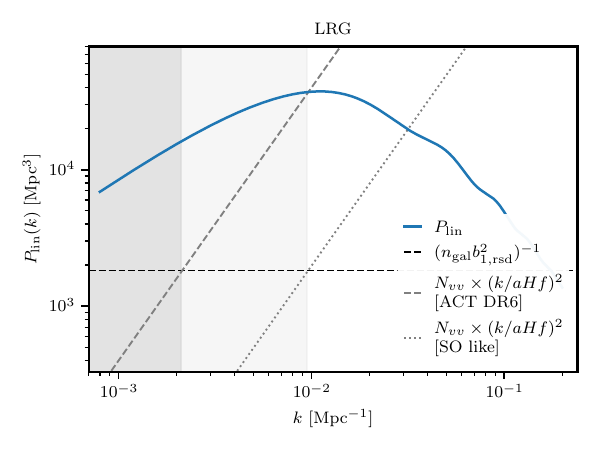}
    \caption{Linear matter power spectrum (blue line) compared to the LRG galaxy shot noise rescaled by the amplitude of the monopole (black dashed line) and to the velocity reconstruction noise $N_{vv}$ for both ACT DR6 measured from this work after correcting for $b_v$  (gray dashed line) and for a realistic SO-like forecast (gray dotted line). The gray regions (dark for the performance with ACT achieved in this work, and light for SO-like forecast) show the scales where the velocity reconstruction from kSZ provides a less noisy measurement.}
    \label{fig:lrg_noise_comparison}
\end{figure}

For the purpose of comparing the performance of our analysis when reconstructing the large-scale density modes, we compare the shot noise intrinsic to galaxy surveys (black dashed line) to the noise on the density field inferred from the velocity reconstruction from ACT DR6 (gray dashed line) performed in this paper. We compare both to the linear matter power spectrum in \cref{fig:lrg_noise_comparison}. This illustrates the constraining power of each methodology over a comparable sky area. In particular, the gray region highlights the scales ($k \lesssim 0.002~\rm{Mpc}^{-1}$) on which velocity reconstruction performs better.

We estimate the velocity noise as $N_{vv} = N_{\widehat{v},\widehat{v}} / b_v^2$, where $N_{\widehat{v},\widehat{v}}$ is estimated directly from the measurement of the velocity–velocity power spectrum (see \cref{fig:LRG_vv}). We then combine the noise contributions from the 90 and 150 GHz channels, including their correlation coefficient.

The gray dotted line shows the forecasted velocity reconstruction noise for a typical SO-like experiment. We start from the map noise level reported in \cite{TheSimonsObservatoryCollaboration2018} and lower the effective noise by $\sim 30\%$ to approximately account for the new “Enhanced” LAT of SO \cite{TheSimonsObservatoryCollaboration2025}, while also incorporating foreground limitations\footnote{The map noise is lowered by a larger amount, but residual foregrounds are lowered at a smaller rate with decrease of CMB map noise.}. In this configuration, velocity reconstruction using the kSZ effect becomes a better alternative for $k \lesssim 0.01~\rm{Mpc}^{-1}$. We note that a similar result holds for both ELGs and QSOs, with comparable scale ranges over which the velocity reconstruction noise falls below the galaxy shot noise.

The results presented here demonstrate the power of this novel methodology for probing the large-scale structure of the Universe, and illustrate the strong potential for cosmic variance cancellation \cite{Seljak2009,Munchmeyer2019,Tishue2025a}. In the coming years, the overlap between CMB and large-scale structure surveys will expand substantially, further enhancing the reach of this approach as a probe of cosmic velocities, large-scale cosmology, and the primordial Universe.
\section{Conclusion} \label{sec:conclusion}
In this work, we present the first three-dimensional kSZ velocity reconstruction using DESI spectroscopic data, based on DESI DR2 in combination with ACT DR6. We report the most significant kSZ detection to date using Luminous Red Galaxies (SNR~$= 17$), as well as the first kSZ detections using Emission Line Galaxies (SNR~$= 8.3$) and quasars (SNR~$= 6.8$), thereby probing the high-redshift Universe out to $z = 3.5$. The combined detection reaches SNR~$=20.8$.

After properly modeling the foreground contributions using the monopole of the galaxy–velocity power spectrum and accurately estimating the velocity reconstruction noise, we achieve the first detection of the velocity–velocity power spectrum using the LRG sample (SNR$= 3.11$). We also note that, although not explored here, the methodology presented in this work could be used to measure the three-dimensional large-scale modes of the linear matter power spectrum through the foregrounds (tSZ effect and CIB).

A remarkable advantage of cross-correlation measurements is that they are largely insensitive to imaging systematics, which can induce spurious excess power on large scales in galaxy–galaxy clustering analyses \cite{Chaussidon2022, Rezaie2023}. The galaxy–velocity power spectra presented in this work follow this expectation: without any dedicated mitigation, the large-scale modes of the galaxy-velocity power spectrum show no anomalous behavior. This robustness allows us to place unbiased constraints on local primordial non-Gaussianity using the velocity field, yielding $\fnl = 15.9_{-34.4}^{+34.6}$ at 68\% confidence.

Despite limitations in the HOD modeling --particularly for ELGs-- and in the electron density profile, our analysis is robust to these choices, as $b_v$ is marginalized over. A more accurate description of the galaxy–electron power spectrum will be explored in future work and is expected to improve the signal-to-noise ratio by more optimally filtering the kSZ information from the CMB temperature maps.

Ultimately, the goal is to combine galaxy–velocity measurements with galaxy–galaxy clustering in order to exploit cosmic variance cancellation \cite{Seljak2009}, providing a powerful avenue for constraining local primordial non-Gaussianity \cite{Munchmeyer2019}. We plan to pursue this approach once the official DESI galaxy–galaxy analysis becomes available. We also note that the extraction of local PNG from large-scale modes can be further optimized by adopting weighting schemes more appropriate than the standard FKP weights used in this work \cite{Castorina2019}, which will be investigated in future work.

Despite the relatively low kSZ response inferred in this work (i.e. low $b_v$), the velocity reconstruction remains a promising method for measuring the distinctive scale-dependent galaxy bias signature in galaxy clustering \cite{Tishue2025a} and for probing inflationary physics in the context of upcoming galaxy surveys such as DESI-II, Euclid \cite{Amendola2018}, LSST \cite{LSSTScienceCollaboration2009}, and SPHEREx \cite{Dore2014}, especially when combined with next-generation CMB observations from the Simons Observatory \cite{TheSimonsObservatoryCollaboration2025}.

We further note that a more detailed investigation of the scale-dependence on small scales ($k>0.04~\rm{Mpc}^{-1}$ \cite{Giri2022}) of the velocity bias $b_v$ would allow the extraction of additional information from the galaxy–velocity power spectrum on these scales. In particular, once $b_v$ is calibrated--either internally using velocity–velocity measurements or externally through observations of Fast Radio Bursts \cite{Madhavacheril2019}--it may become possible to infer the growth rate of structure with high precision in combination with the galaxy-galaxy power spectrum \cite{Wang2025, Qin2025}.
\section*{Data availability}
Data points of each plot are made available on Zenodo\footnote{\url{https://zenodo.org/records/21938200}} as part of DESI’s Data Management Plan. DESI DR2 data have not been publicly released yet.

\begin{acknowledgments} 
We thank Alina Sabyr and Yulin Gong for agreeing to act as DESI internal reviewers. E.C. and S.F. are supported by Lawrence Berkeley National Laboratory and the Director, Office of Science, Office of High Energy Physics of the U.S. Department of Energy under Contract No.\ DE-AC02-05CH11231. S.C.H. is supported by the P.J.E.~Peebles Fellowship at the Perimeter Institute. Research at Perimeter Institute is supported in part by the Government of Canada through the Department of Innovation, Science and Economic Development and by the Province of Ontario through the Ministry of Colleges and Universities. X.C.\ is supported by the U.S.\ Department of Energy.

This material is based upon work supported by the U.S. Department of Energy (DOE), Office of Science, Office of High-Energy Physics, under Contract No. DE–AC02–05CH11231, and by the National Energy Research Scientific Computing Center, a DOE Office of Science User Facility under the same contract. Additional support for DESI was provided by the U.S. National Science Foundation (NSF), Division of Astronomical Sciences under Contract No. AST-0950945 to the NSF’s National Optical-Infrared Astronomy Research Laboratory; the Science and Technology Facilities Council of the United Kingdom; the Gordon and Betty Moore Foundation; the Heising-Simons Foundation; the French Alternative Energies and Atomic Energy Commission (CEA); the National Council of Humanities, Science and Technology of Mexico (CONAHCYT); the Ministry of Science, Innovation and Universities of Spain (MICIU/AEI/10.13039/501100011033), and by the DESI Member Institutions: \url{https://www.desi.lbl.gov/collaborating-institutions}. Any opinions, findings, and conclusions or recommendations expressed in this material are those of the author(s) and do not necessarily reflect the views of the U. S. National Science Foundation, the U. S. Department of Energy, or any of the listed funding agencies.

The authors are honored to be permitted to conduct scientific research on I'oligam Du'ag (Kitt Peak), a mountain with particular significance to the Tohono O’odham Nation.

We acknowledge the use of public ACT data products made available through the National Energy Research Scientific Computing Center (NERSC), a U.S. Department of Energy Office of Science User Facility operated under Contract No. DE-AC02-05CH11231.
\end{acknowledgments}

\bibliographystyle{bibli_style}
\bibliography{bibli}
\appendix
\section{Halo model for foregrounds} \label{app:foreground_model}

In this appendix, we'll argue that on large scales, CMB foregrounds produce an extra term $(b_{\rm fg} \delta)$ in the velocity reconstruction, of the form:
\begin{equation}
\big\langle \widehat{v}_r(\vb{x}) \big\rangle = b_v v_r^{\rm true}(\vb{x}) + b_{\rm fg} \delta(\vb{x}) + \mbox{noise} \,.
\label{eq:fg_line1}
\end{equation}
We will show this by brute-force calculation in the halo model.

It suffices to consider a single frequency channel, and a single CMB foreground (e.g.\ tSZ or CIB).

For simplicity, we use a ``snapshot'' geometry (following \cite{Smith2018}), where the universe is a 3-d periodic box at a fixed time, and the CMB is a 2-d flat-sky field along one face of the box.
The snapshot geometry is an approximation of the true ``lightcone'' geometry, and neglects sky curvature and redshift evolution.

In the snapshot geometry, slowly evolving quantities, such as $\chi(z)$ or $K(z)$, are evaluated at a fixed time, and denoted $\chi_*$ or $K_*$.
We identify 2-d positions $\boldsymbol{\theta}$ with transverse 3-d positions $\vb{x}_\perp = \chi_* \boldsymbol{\theta}$, and identify 2-d wavenumbers $\vb{l}$ with transverse 3-d wavenumbers $\vb{k} = (\vb{l}/\chi_*)$.

We model foreground emission as a sum over halos:
\begin{equation}
T_{\rm fg}(\vb{l}) = \sum_j T_j \, u_l(M_j) \, e^{i\vb{l}\cdot\boldsymbol{\theta}_j} \label{eq:tfg}
\end{equation}
where $T_j$ is the total foreground flux from the $j$-th halo, and $u_l(M)$ is an angular profile normalized so that $u_l(M)=1$ at $l=0$.
Similarly, we model the galaxy field as a sum over halos:
\begin{equation}
\rho_g(\vb{x}) = \sum_i N_i \delta^3(\vb{x}-\vb{x}_i)
\end{equation}
where $N_i$ is 0 or 1 depending on whether the $i$-th halo contains a galaxy.
For simplicity, we have assumed that all galaxies are centrals.

In the snapshot geometry, the kSZ velocity reconstruction is defined by $\widehat{v}_r = \rho_g \cdot \widetilde{T}$, where $\widetilde{T}$ is the high-pass filtered CMB:
\begin{equation} \label{eq:hvr_def}
\begin{cases}
\widehat{v}_r(\vb{x}) &\equiv \sum_i N_i \, \widetilde{T}(\boldsymbol{\theta}_i) \, \delta^3(\vb{x}-\vb{x}_i) \\[1ex]
\widetilde{T}(\vb{l}) &\equiv F_l \, T(\vb{l})
\end{cases}  \,,
\end{equation}
where $F_l$ is a CMB filter which is zero on large scales ($l \le l_{\rm min}^{\rm ksz} \sim 2000$).

Plugging Eq.\ (\ref{eq:tfg}) into (\ref{eq:hvr_def}), the foreground contribution to $\widehat{v}_r(\vb{x})$ can be written as a double sum over halos:
\begin{equation}
\widehat{v}_r(\vb{x}) = \sum_{ij} N_i T_j \, \delta^3(\vb{x}-\vb{x}_i) \int_{\vb{l}} F_l \, u_l(M_j) \, e^{i\vb{l}\cdot(\boldsymbol{\theta}_i-\boldsymbol{\theta}_j)} \,.
\label{eq:fg_double_sum}
\end{equation}
This expression will be our starting point for computing the expectation value $\langle \widehat{v}_r(\vb{x}) \rangle$.

Recall that in the halo model, large-scale structure is modeled as a two-step random process: first, we sample a random {\em linear} density field $\delta(\vb{x})$, then we randomly place Poisson halos.
When we compute $\langle \widehat{v}_r(\vb{x}) \rangle$ in the halo model, we will take the conditional expectation value over Poisson halo placements, in a fixed realization of $\delta(\vb{x})$.
Thus, the expectation value $\langle \widehat{v}_r(\vb{x}) \rangle$ will depend on the realization $\delta(\vb{x})$, as in Eq.\ (\ref{eq:fg_line1}).

The expectation value of the double sum (\ref{eq:fg_double_sum}) consists of 1-halo and 2-halo terms.
The 1-halo term is straightforward to compute:
\begin{equation}
\begin{aligned}
\big\langle \widehat{v}_r(\vb{x}) \big\rangle_{1h}
 &= \int dM \, n(M) \big[ 1 + b(M) \delta(\vb{x}) \big] \big\langle NT \big\rangle_M \\
 & \hspace{1.5cm} \times 
    \int \frac{d^2\vb{l}}{(2\pi)^2} F_l \, u_l(M)
\end{aligned}
\end{equation}
where we have assumed that the expectation value $\langle N_i T_i \rangle$ only depends on halo mass $M_i$, and denoted it by $\langle N T \rangle_M$.
We drop the ``DC'' term on the RHS, which is constant in $\vb{x}$, and keep the term proportional to $\delta(\vb{x})$. In our pipeline, the DC term is projected out by the mean-subtraction in Eq.\ (\ref{eqn:velocity_estimator}).
The result can be written:
\begin{equation}
\big\langle \widehat{v}_r(\vb{x}) \big\rangle_{1h} = b_{\rm fg} \delta(\vb{x})
\end{equation}
where $b_{\rm fg}$ is given by:
\begin{equation} \label{eqn:bfg_theo}
b_{\rm fg} \equiv \int dM \, n(M) b(M) \, \big\langle NT \rangle_M 
\int \frac{d^2\vb{l}}{(2\pi)^2} F_l \, u_l(M) \,.
\end{equation}
Next, consider the two-halo term.
Starting from the double sum (\ref{eq:fg_double_sum}), we get:

\begin{widetext}
\begin{equation}
\big\langle \widehat{v}_r(\vb{x}) \big\rangle_{2h} = \int d^3\vb{x}' \, dM \, dM' \, n(M) n(M') \, \langle N \rangle_M \, \langle T \rangle_{M'} \times \left[ 1 + b(M) \delta(\vb{x}) \right] \left[ 1 + b(M') \delta(\vb{x}') \right] \times \int \frac{d^2\vb{l}}{(2\pi)^2} F_l \, u_l(M') e^{i\vb{l}\cdot(\boldsymbol{\theta}-\boldsymbol{\theta}')} \,.
\end{equation}
\end{widetext}

We Fourier transform $\vb{x} \rightarrow \vb{k}_L$ where $\vb{k}_L$ is a large scale.
Since $F_l$ is zero on large scales $l \sim (k_L/\chi_*)$, only the term proportional to $\delta(\vb{x}) \delta(\vb{x}')$ survives.
After a little algebra, it can be written:
\begin{equation}
\big\langle \widehat{v}_r(\vb{k}_L) \big\rangle_{2h}
 = \int \frac{d^2\vb{l}}{(2\pi)^2} A_l \Big[ \delta(\vb{l}/\chi_*) \, \delta(\vb{k}_S) \Big]_{\vb{k}_S=\vb{k}_L-\vb{l}/\chi_*}
 \label{eq:vr_2h}
\end{equation}
where $A_l$ is defined by:
\begin{equation}
\begin{aligned}
A_l &\equiv F_l \int dM \, dM' \, n(M) n(M') \, b(M) b(M') \\
 & \hspace{1cm} \times
   \big\langle N \big\rangle_M \, 
   \big\langle T \big\rangle_{M'} \,
   u_l(M') \,.
\end{aligned}
\end{equation}
The RHS of Eq.\ (\ref{eq:vr_2h}) is uncorrelated to the large-scale fields $\delta(\vb{k}_L)$ and $v_r(\vb{k}_L)$, since it only involves small-scale modes of the field $\delta(\vb{x})$, which is Gaussian in the halo model.
It can therefore be interpreted as a small contribution to the reconstruction noise.
Note that the RHS of Eq.\ (\ref{eq:vr_2h}) has a power spectrum which is constant in $k_L$, on large scales $k_L \ll (l_{\rm min}^{\rm ksz}/\chi_*)$.

Summarizing this section, we have shown that each foreground (tSZ, CIB, etc.) produces a term in the velocity reconstruction of the form:
\begin{equation}
\big\langle \widehat{v}_r(\vb{x}) \big\rangle = \underbrace{b_{\rm fg} \delta(\vb{x})}_{\rm 1-halo} + \underbrace{\mbox{noise}}_{\rm 2-halo} \,.
\end{equation}

\begin{figure*}[!hp]
    \centering
    \vspace{2cm}
    \includegraphics[scale=0.9]{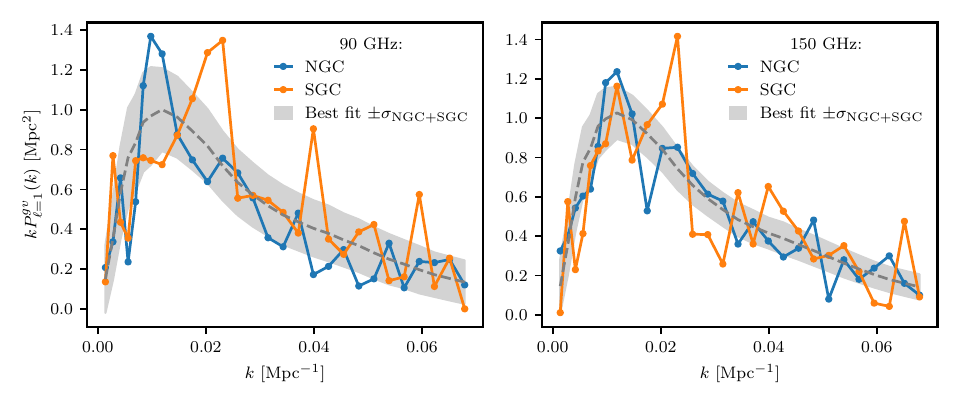}
    \includegraphics[scale=0.9]{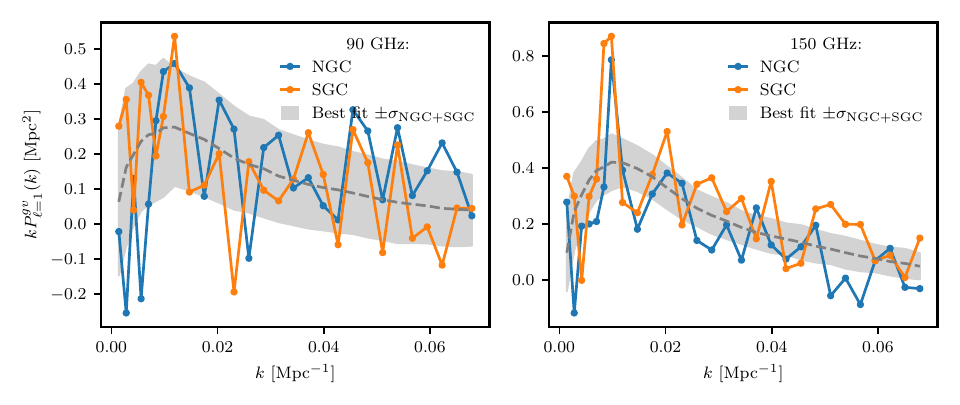}
    \includegraphics[scale=0.9]{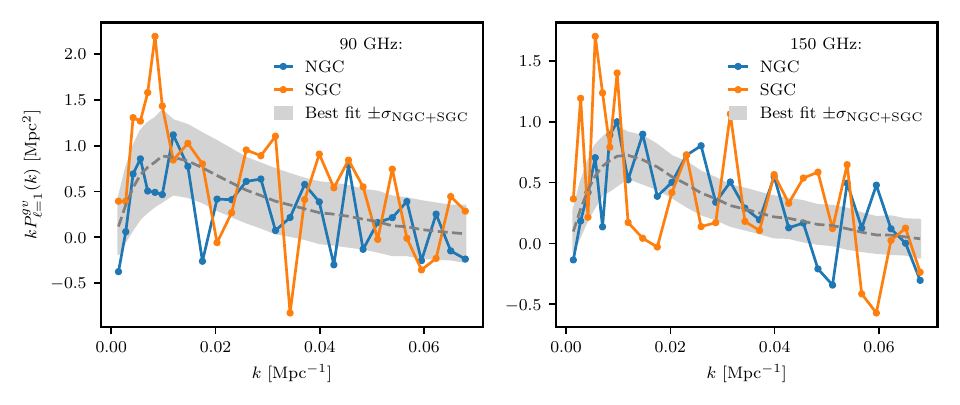}
    \caption{Comparison between the NGC (blue) and SGC (orange) dipole of galaxy-velocity power spectrum for the LRG, ELG and QSO (from top to bottom) sample. Grey shaded regions are the $1\sigma$ errors from the combination of the two drawn around the best fit.}
    \vspace{2cm}
    \label{fig:ngc_vs_sgc}
\end{figure*}

\section{NGC vs. SGC} \label{app:ngc_vs_sgc}
In order to test the reliability of our measurement, we compare the dipole of the galaxy-velocity power spectrum computed in the North Galactic Cap (NGC) and in the South Galactic Cap (SGC). These two regions are independent and can then be compared directly. This comparison is done in \cref{fig:ngc_vs_sgc} for LRGs (top), ELGs (middle), and QSOs (bottom). Since the gray regions are $1 \sigma$ errors drawn around the best fit for the combination of NGC and SGC, as explained in \cref{sec:data}, we do not observe any individual point that highlights a tension. 

\begin{figure}[h!]
    \centering    
    \includegraphics[width=\linewidth]{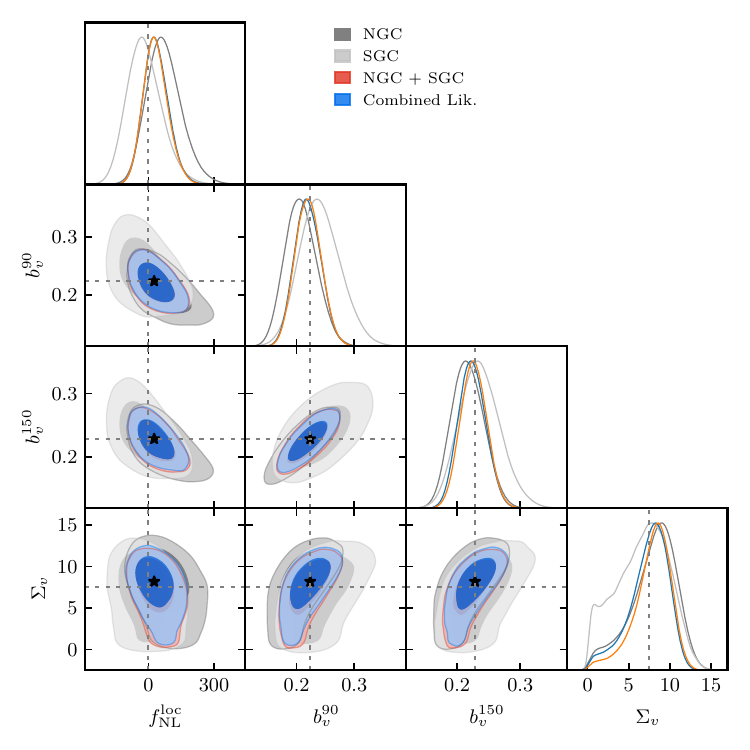}
    \caption{Posterior comparison for the LRG sample between the inference done in the NGC (dark gray), SGC (light gray) and the combination of the two either at the data vector level (red) or at the likelihood level (blue). The fiducial analysis choice is to combine at the data vector level.}
    \label{fig:lrg_ngc_vs_sgc}
\end{figure}

\cref{fig:lrg_ngc_vs_sgc} compares the inferences for the LRG sample done individually in NGC and SGC to the one from the combined measurement, either at the data vector level (fiducial choice) or at the likelihood level. Again, the NGC and SGC are perfectly consistent. Similar results hold for ELGs and QSOs, although they are noisier.

\section{Null test (150-90)} \label{app:null_test}
\begin{figure}
    \centering
    \includegraphics[width=\linewidth]{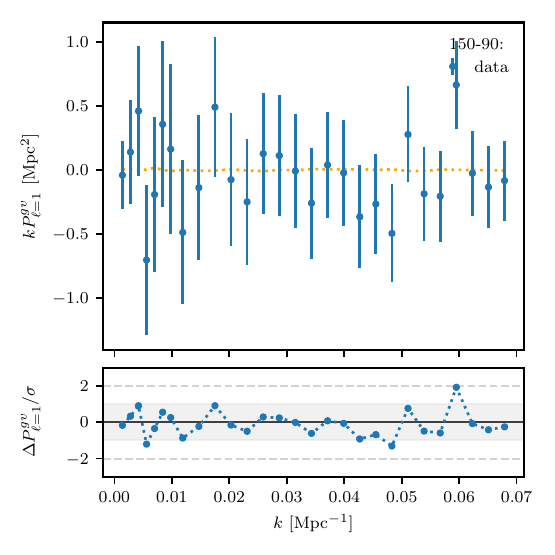}
    \caption{Null test. LRG dipole of the galaxy-velocity power spectrum where the velocity is reconstructed as the difference of $\widehat{v}_r^{150} - \widehat{v}_r^{90}$.}
    \label{fig:lrg_null_test}
\end{figure}

Since the two velocity reconstructions are probing the same underlying velocity field, the difference $\widehat{v}_r^{150} - \widehat{v}_r^{90}$ should not contain any kSZ information. Hence, measuring $P_{\ell=1}^{g v^{150-90}}$ can be viewed as a null test. This dipole is displayed for the LRG case in \cref{fig:lrg_null_test} and it does not show any deviation to zero.

\section{Octopole of the Galaxy-Velocity Power Spectrum} \label{app:octopole}
Similarly to the galaxy-galaxy power spectrum, we try to detect higher order multipoles of the galaxy-velocity power spectrum, namely the octopole ($\ell=3$). Because the velocity bias $b_v$ is treated as a nuisance parameter, the octopole could provide an independent way to calibrate it:
\begin{equation} \label{eqn:pgv_ell3_theo}
     P_{\ell=3}^{gv} = \dfrac{2}{5} b_v \dfrac{if^2aH}{k} P_{\rm lin}(k) \,.
\end{equation}
Note that with the definition of \cref{eqn:spin-l}, the estimated $\widehat{P}_{\ell}^{gv}$ is real and does not have the $i$ in \cref{eqn:pgv_ell3_theo}. 

\begin{figure*}[!ht]
    \centering
    \includegraphics[scale=0.9]{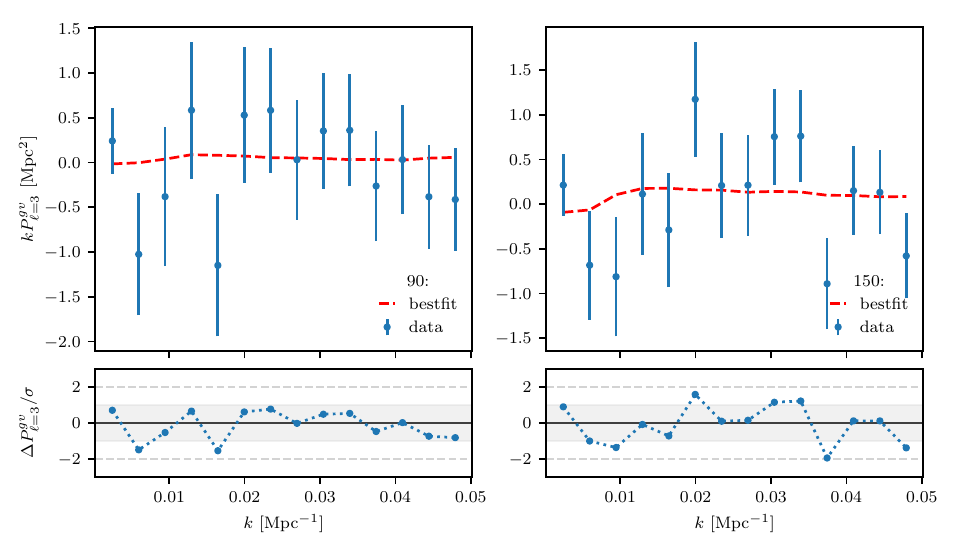}
    \caption{Similar to \cref{fig:LRG_gv} but for the octopole ($\ell=3$) of the LRG galaxy-velocity power spectrum, which is not detected (SNR $\sim 0.75$).}
    \label{fig:LRG_gv_ell3}
\end{figure*}

\begin{figure}[!h]
    \centering
    \includegraphics[scale=0.9]{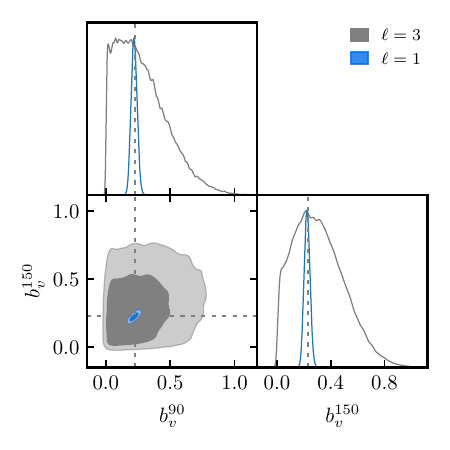}
    \caption{Comparison between the posteriors from the dipole ($\ell=1$ in blue) and from the octopole ($\ell=3$ in gray) of the LRG galaxy-velocity power spectrum. Best fit values are given in \cref{tab:lrg_gv_ell3}.}
    \label{fig:LRG_mcmc_gv_ell3}
\end{figure}

The measured octopole of the galaxy-velocity power spectrum of the LRG sample is displayed in \cref{fig:LRG_gv_ell3}. We have rebin the data vector by a factor 2 in order to reduce the noise in each bin. We tried to measure the amplitude of the signal as shown in \cref{fig:LRG_mcmc_gv_ell3} with the best fit values given in \cref{tab:lrg_gv_ell3}. However, this measurement has a very low SNR ($\sim 0.74$). 

There are two remarks that can be made here. First, these measured octopoles do not exhibit any large-scale power, illustrating again the insensitivity of the cross-correlation to imaging systematics (quadrupole is also highly contaminated by these systematics in the case of galaxy-galaxy). Then, even if the error bars are quite large, the amplitudes of the signal is well consistent with the amplitude inferred from the dipole as shown in \cref{fig:LRG_mcmc_gv_ell3}. 

This signal will be detected only with both smaller statistical noise and velocity reconstructed noise.

\begin{table}[!h]
    \centering
    \caption{Best-fit results for the octopole of the LRG galaxy-velocity power spectra with SNR~$=0.74$ and $\chi^2=21.7$ with 26 degrees of freedom leading to $\chi_{red}^2=0.834$ and PTE~$=0.71$.}
    \vspace{1.5mm}
    \begin{tabular}{lccc}
    \toprule
     Params     &   BestFit &   Mean & $\pm \sigma$ \\ \midrule
    $b_v^{90}$  &     0.104 & 0.28 & -0.199/0.198 \\
    $b_v^{150}$ &     0.21 &  0.31 & -0.193/0.192 \\
    \bottomrule
    \end{tabular}
    \label{tab:lrg_gv_ell3}
\end{table}

\section{$\widehat{P}_{\ell=0,\ell=0}^{vv}$ versus $\widehat{P}_{\ell=1,\ell=1}^{vv}$} \label{app:ell00vsell11}
As introduced in \cref{eqn:pvv_estimator}, we can estimate the velocity-velocity power spectrum with different approaches. In particular, we can weigh the velocity field with different $\mu$ power, leading to two different estimators: $\widehat{P}_{\ell=0,\ell=0}^{vv}$, $\widehat{P}_{\ell=1,\ell=1}^{vv}$.

The $\mu$ weighting is nulling some foreground contributions at large scales in the same way as for dipole, with only a remaining contamination from the leakage due to the window function. In reality, it is not a problem since we are able to correctly model this effect and derive the amount of foregrounds from the monopole of the galaxy-velocity power spectrum. However, this weighting is the optimal way to extract the $\mu$ dependence from the velocity field as shown in \cite{Tegmark1997} and, for this reason, we are using it as the default in our analysis.

\begin{table}[!h]
    \centering
    \caption{Best-fit results for the LRG velocity-velocity power spectra (including the cross-correlation) for $\mu^0$ weighting scheme (top) with SNR~$=1.51$ and $\chi^2=40.90$ with 57 degrees of freedom leading to $\chi_{red}^2=0.718$ and PTE~$=0.94$ and $\mu^1$ weighting scheme (bottom) with SNR~$=3.11$ and $\chi^2=51.12$ with 57 degrees of freedom leading to $\chi_{red}^2=0.897$ and PTE~$=0.69$. Both SNRs are given after noise and foregrounds subtraction.}
    \vspace{1.5mm}
    \begin{tabular}{lccc}
    \toprule
     Params           & BestFit & Mean  & $\pm \sigma$     \\
     \midrule \multicolumn{4}{c}{$P_{\ell=0,\ell^\prime=0}^{vv}$} \\ \midrule
     $b_v^{90}$       & 0.24   & 0.189  & -0.115/0.106  \\
     $b_v^{150}$      &  0.173 & 0.156 & -0.0931/0.0858   \\
     $\Sigma_v^{vv}$  & 8.32e-05    & 23.6    & -16.2/16.4     \\
     $s_{nv}^{90}$    & 1.13    & 1.13  & -0.00625/0.00625 \\
     $s_{nv}^{150}$   & 1.1    & 1.1  & -0.00460/0.00459 \\
     $s_{nv}^{90x150}$& 1.29    & 1.29  & -0.0141/0.0141   \\
     \midrule \multicolumn{4}{c}{$P_{\ell=1,\ell^\prime=1}^{vv}$} \\ \midrule
     $b_v^{90}$       & 0.298   & 0.273 & -0.0786/0.0778   \\
     $b_v^{150}$      & 0.248   & 0.232 & -0.0639/0.0643   \\
     $\Sigma_v^{vv}$  & 12.3    & 21.0  & -14.6/15.0       \\
     $s_{nv}^{90}$    & 1.13    & 1.13  & -0.00705/0.00703 \\
     $s_{nv}^{150}$   & 1.09    & 1.09   & -0.00559/0.00561 \\
     $s_{nv}^{90x150}$& 1.28    & 1.29  & -0.0172/0.0172   \\ 
    \bottomrule
    \end{tabular}
    \label{tab:LRG_vv}
\end{table}
 
This gain of information is illustrated in \cref{tab:LRG_vv}. Here, we fix the foreground's contribution to its best-fit values, see \cref{sec:lrg_foregrounds}. The constraining power on $b_v$ --\textit{i.e.} detecting the velocity part of the signal-- from $P_{\ell=0,\ell^\prime=0}^{vv}$ is weaker with SNR~$\sim 1.51$ than from $P_{\ell=1,\ell^\prime=1}^{vv}$ with SNR~$\sim 3.11$. Both are given consistent measurements of the amplitude of the velocity-velocity signal. Note that, as expected, since the velocity reconstruction noise has no $\mu$ dependence, it is less constraining using $P_{\ell=1,\ell^\prime=1}^{vv}$ than with $P_{\ell=0,\ell^\prime=0}^{vv}$.

\section{Additional figures}
In this section, we are describing some additional materials that are referenced as an appendix in the main text. 

\begin{figure*}
    \centering
    \vspace{1.5cm}
\includegraphics[width=\textwidth]{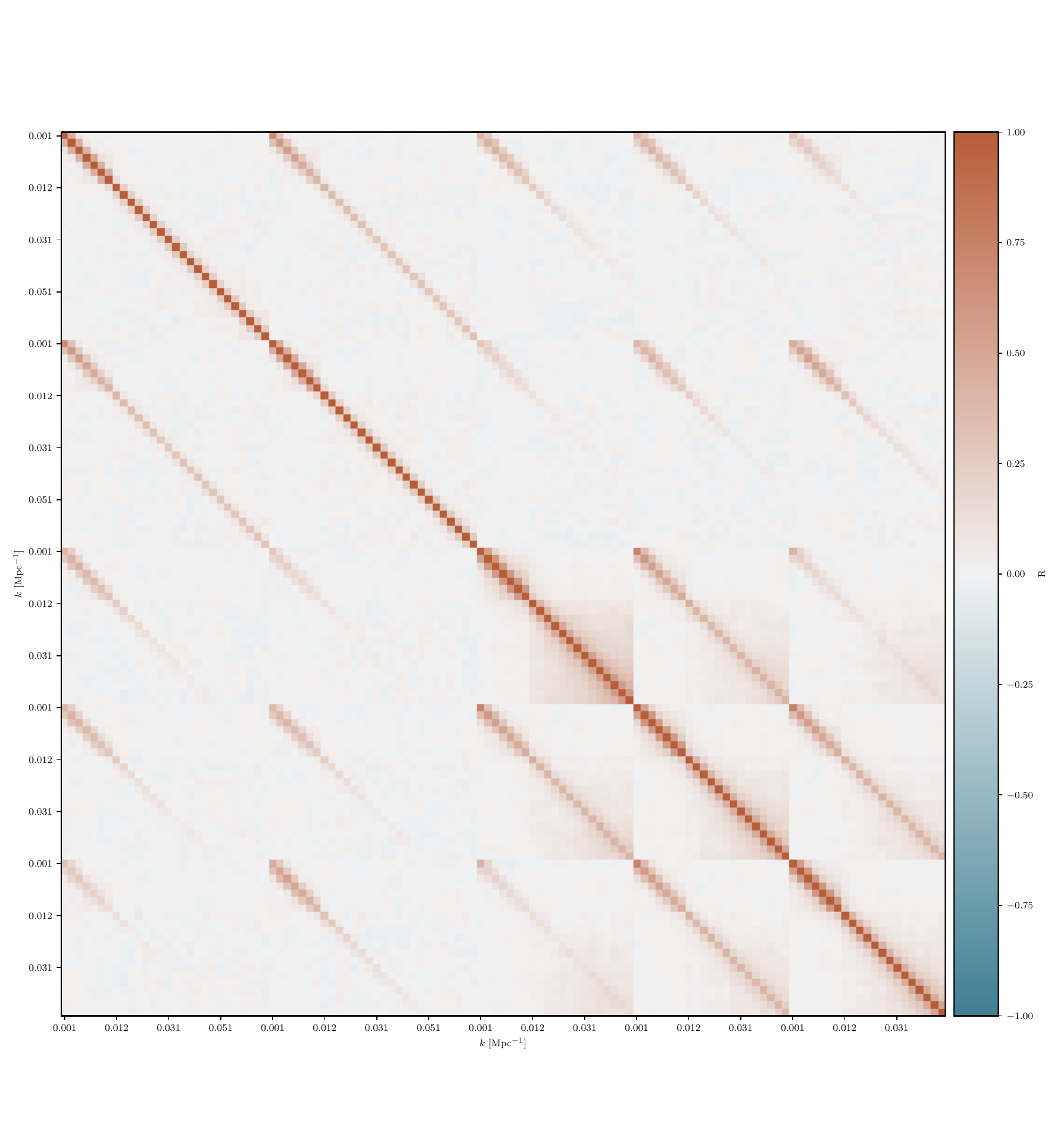}
    \caption{Correlation matrix for the LRG joint analysis. The observables are stacking in the following order: $g$~x~$v^{90}$, $g$~x~$v^{150}$, $v^{90}$~x~$v^{90}$, $v^{90}$~x~$v^{150}$, $v^{150}$~x~$v^{150}$.}
    \label{fig:corr_example}
    \vspace{2cm}
\end{figure*}

First, \cref{fig:corr_example} shows the correlation matrix, estimated from our surrogates methodology presented in \cref{sec:surrogates}, for the LRG joint analysis of the galaxy-velocity power spectra and the velocity-velocity power spectra, \cref{sec:lrg_analysis}. The observables are stacked in the following order: $g$~x~$v^{90}$, $g$~x~$v^{150}$, $v^{90}$~x~$v^{90}$, $v^{90}$~x~$v^{150}$, $v^{150}$~x~$v^{150}$.

Several features can be clearly distinguished in this correlation matrix. The two galaxy-velocity power spectra are indeed correlated since they are tracing the same underlying velocity field. The correlation is larger (up to 0.8) at large scales where the velocity reconstruction noise is smaller. 

A similar comment can be made on the velocity-velocity part of the correlation matrix and in particular in the $v^{90}$~x~$v^{90}$ -- $v^{150}$~x~$v^{150}$ block where at small scales the two signals, dominated by noise, are not correlated at all. We also note that for $v^{90}$~x~$v^{150}$ the small scales in the diagonal highlight common noise from similar primary CMB in the 90 and 150 GHz channels.
 
\begin{figure*}
    \centering
    \vspace{2.5cm}
    \includegraphics[width=1\linewidth]{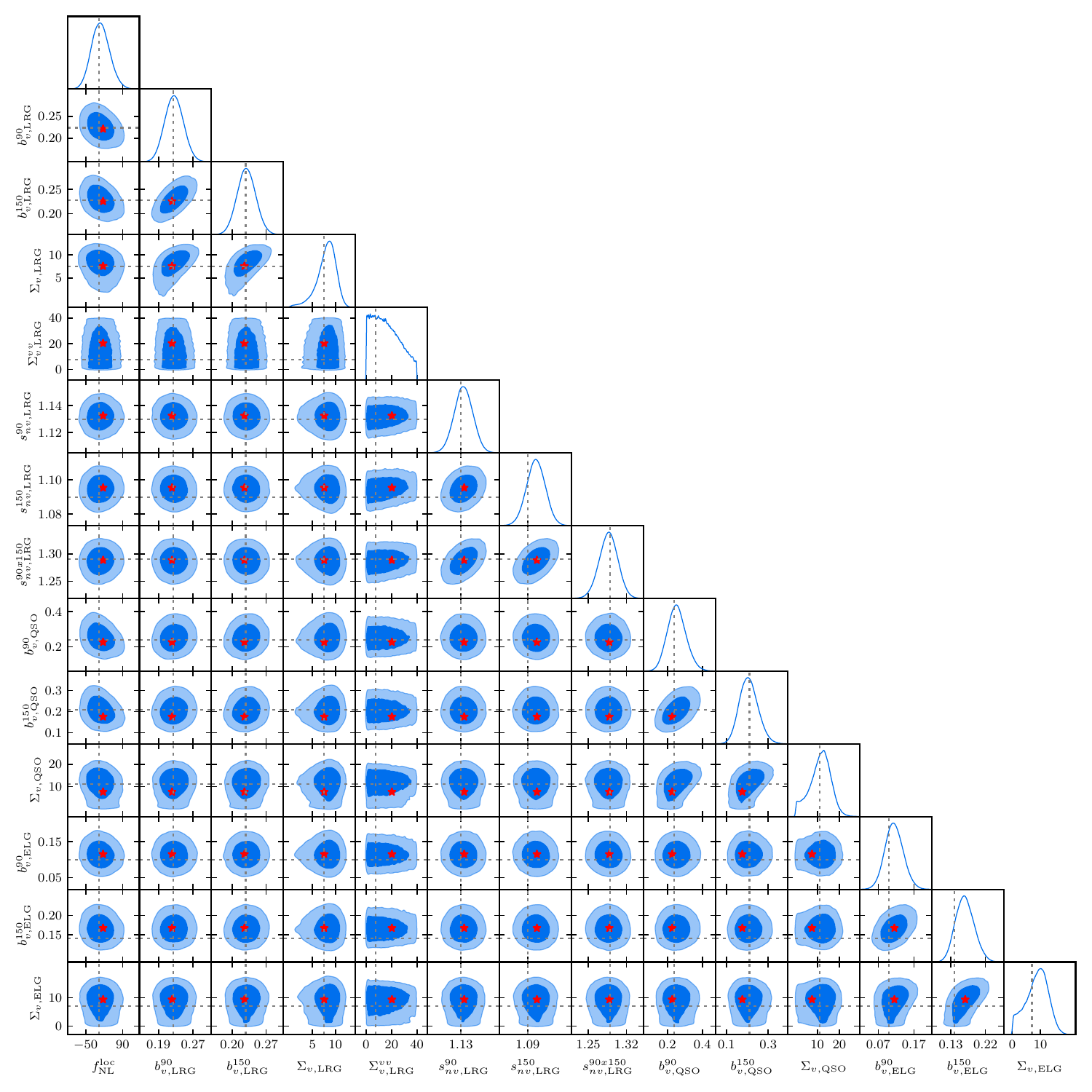}
    \caption{Full posteriors for the combined fit of the LRG, ELG, and QSO. The gray dashed lines correspond to the fiducial parameter values adopted for the covariance, while the red stars are the best-fit values from the profiling. As expected, $\Sigma_{v,\rm{LRG}}^{vv}$ is poorly constrained, see \cref{sec:lrg_analysis}.}
    \label{fig:combined_tracers_full}
    \vspace{2cm}
\end{figure*}

The last figure, \cref{fig:combined_tracers_full}, of this appendix is the full posteriors of the combination of the LRG, ELG and QSO presented in \cref{sec:full_analysis}.

\end{document}